
\magnification 1000
\baselineskip 12pt
\overfullrule=0pt

\def\newpage{\vfill\eject}
\def\newline{\hfill\break}
\def\Det{{\rm Det}}
\def\val{{\rm val}}
\def\S{{\cal S}}
\def\P{{\cal P}}
\def\A{{\cal A}}
\def\V{{\cal V}}
\def\F{{\cal F}}

\def\L{{\cal L}}
\def\M{{\cal M}}
\def\D{{\bf D}}
\def\tr{\,{\rm tr}\,}
\def\({\Bigl(}
\def\){\Bigr)}
\def\]{\Bigr]}
\def\[{\Bigl[}

\def\slashchar#1{\setbox0=\hbox{$#1$}
   \dimen0=\wd0 \setbox1=\hbox{/} \dimen1=\wd1
   \ifdim\dimen0>\dimen1 \rlap{\hbox to \dimen0{\hfil/\hfil}} #1
   \else  \rlap{\hbox to \dimen1{\hfil$#1$\hfil}} / \fi}


\null
\centerline{\bf CHIRAL ANOMALY AND NUCLEON PROPERTIES }
\centerline{\bf IN THE  NAMBU--JONA-LASINIO MODEL WITH VECTOR MESONS}
\vskip1cm
\centerline{ E. Ruiz Arriola ${}^{1,2}$ and L.L. Salcedo ${}^1$ }
\vskip0.5cm
\centerline {${}^1$Departamento de F\'{\i}sica Moderna, Universidad de Granada}
\centerline {E-18071 Granada, Spain }
\vskip0.5cm
\centerline{ ${}^2$National Institute for Nuclear Physics and High Energy
Physics, (NIKHEF-K)}
\centerline{ 1009-DB Amsterdam, The Netherlands}
\vskip1cm
\centerline{\bf ABSTRACT}
\vskip 18pt
\vbox{
\baselineskip 12pt
\narrower{
We consider the extended SU(3) Nambu and Jona-Lasinio model with
explicit vector couplings in the presence of external fields.
We study the chiral anomaly in this model and its implications on the
properties of the nucleon described as a chiral soliton of three
valence quarks bounded in mesonic background fields. For the model to
reproduce the QCD anomaly it is necessary to subtract suitable local
and polynomial counterterms in the external and dynamical vector and
axial-vector fields. We compute the counterterms explicitly in a vector
gauge invariant regularization, and obtain modifications to the total
effective action and vector and axial currents. We study the
numerical influence of those counterterms in the two flavour version of
model with dynamical $\sigma,\pi, \rho, A $ and $\omega$ mesons. We
find that, for time independent hedgehog configurations, the numerical
effects in the nucleon mass, the isoscalar nucleon radius and the axial
coupling constant are negligibly small.
}}
\vskip1cm
\centerline{\sl 17 March 1995 ( revised) }
\centerline{\sl Submitted to Nuclear Physics {\bf A} }
\vfill
\line{UG-DFM-1/95 \hfill}

\newpage
\parskip.5cm

{\bf 1. Introduction}

The QCD chiral anomaly in its vector gauge invariant form
[1] represents an important constraint for
any effective low energy model of hadronic interactions. It is
responsible for the $\pi^0 \to 2 \gamma $ decay, provides evidence
that the number of colours $N_c$ is equal to three and is an exact one
loop result not subjected to renormalization [2]. From the point of
view of the effective Lagrangian and depending upon the degrees of
freedom involved, there are at least two ways in which the chiral
anomaly can be introduced. In a pure mesonic theory
one can add the vector gauged Wess-Zumino term  [3,4] to the
action
which by definition saturates the QCD anomaly equation. This type of
construction and similar ones have been extensively used to calculate
abnormal parity mesonic processes [5,6] or to provide short
range repulsion in the construction of topological chiral soliton
models  of the Skyrme type [7]. Proceeding in this way there is in
principle an ambiguity
concerning the possible abnormal parity non-anomalous terms which do
not contribute to the anomaly equation. One should say, however, that
the ambiguity is completely removed if only pions and external fields
are considered, but remains if other particles like e.g. heavy vector
resonances are included. This allows to fix the
corresponding coefficients in the chiral Lagrangian to fulfill some
desired properties motivated by phenomenology and not directly
derivable from the chiral anomaly [6].

In a relativistic chiral quark model the
situation is slightly different. The non invariance of the functional
integration measure or equivalently of the fermion determinant under
chiral transformations [8,9] guarantees the onset of a chiral
anomaly
in the effective theory whose particular form depends on the symmetries
preserved by the regulator. Actually, it is by no means clear that the
resulting anomaly coincides with the QCD anomaly. In such a situation,
the question arises whether the quark model can be redefined in such a
way that the correct QCD anomaly can be reproduced.

A prototype of a chiral quark model for hadronic structure is represented
by the Nambu--Jona-Lasinio (NJL) model [10]. This model has been
studied intensively in the
vacuum sector, the meson sector and the baryon sector [11]. It is a
pure quark model which incorporates explicit, spontaneous and
anomalous chiral symmetry breaking, describes mesons as quark-antiquark
pairs and baryons as solitons of three bound quarks or alternatively as
quark-diquark bound states. However, it does not include confinement
and it is not renormalizable requiring the use of a low energy cut-off.

The generalization of the model to include vector and axial-vector
couplings has also been studied in much detail either in a bosonized
version [12,13,14,15,16,17] or in a pure quark language
[18,19,20,21] since it allows to describe more mesonic
phenomenology
and provides a realization of vector meson dominance through current
field identities [22].

Actually, we have shown in a recent paper [23] that the Nambu and
Jona-Lasinio model with
scalar, pseudoscalar, vector and axial mesons as it is most often used
does not reproduce the anomaly of QCD, the correct result being
reproduced if the vector couplings are set
equal to zero. This result is perfectly consistent with vector gauge
invariance, i.e. only the divergence of the axial current is modified
with respect to the QCD result whereas the vector current is still
conserved. Curiously, the $\pi^0 \to 2\gamma $ decay amplitude
remained unchanged, but other anomalous amplitudes
involving more than one pion, as for instance $\gamma \to 3\pi $
deviated from the current algebra result by  $ \sim 20\% $ for typical
values of the parameters. This is an unpleasant feature  because
questions the meaning of previous calculations involving anomalous
vertices, both in the meson [24] and the soliton sector
[25,26,27,28,29,30,31] based on actions with vector
mesons not containing the proper anomalous
structure [1]. The problem arises whether the non fulfillment of the
QCD chiral anomaly is a problem of describing vector mesons through
vector couplings or whether it can be mended by addition  of legitimate
counterterms in order to reproduce the QCD
anomaly in its Bardeen form. In any case, it is desirable to know
the possible implications  not only for mesonic decays but also
at the nucleon level. To overcome this problem Bijnens and Prades [32]
have suggested, following elegant mathematical arguments already given
by Bardeen and Zumino [33] and pursued by others (see e.g. [5] and
references included in [32]), that
indeed
such counterterms exist and that they are unique if CP invariance is
invoked. This represents another  independent non trivial solution
to the anomaly equation not considered explicitly in our previous
work [23]. The choice between these two solutions
depends on our desire to include explicit vector and axial
couplings in the starting NJL Lagrangian and at the same time being able
to reproduce the QCD chiral anomaly. However, it should
be stressed that the subtraction involves not only the external fields
but also the dynamical vector fields. From the point of view of
perturbation theory this corresponds to change an infinite number of
diagrams in contrast to the usual external field subtraction procedure.
In this respect this is a very particular feature of the NJL model and
its singularity structure.
As a consequence vector meson dominance in the usual sense is lost.
This means that a consistent Lagrangian incorporating both the QCD
anomaly and complete vector meson dominance cannot be constructed,
at least in the vector field representation of vector mesons.

In the approach of ref. [32] the calculation of the
counterterms requires an explicit knowledge of the vector and
axial currents. In addition, the formulas given by those
authors are rather compact and of low practical utility in the present
context. In the present paper we adhere to the point of view of
ref. [32] although
propose a somewhat different methodology which only relies on the
knowledge of the anomaly itself. In fact we trace back the origin of the
ambiguity to the natural regularization suggested by the bosonization
procedure widely used for performing low energy expansions,
determination of mesonic vertex
functions and description of baryons as solitons. We derive the
modifications of the effective action and the vector and axial currents
and obtain the leading large $N_c$ corrections to the Current-Field
Identities [22]. We also particularize the resulting formulas for the
the specific two flavour case with $\sigma$, $\pi$, $\rho $, $A$ and
$\omega$ mesons and compute the numerical modification of the nucleon
energy, the isoscalar nucleon radius and the axial coupling constant
in the solitonic picture of baryons. As a byproduct we also compute the
leading low energy chiral invariant, i.e. non anomalous,  contribution
to the abnormal parity action of vector mesons in the presence of
external fields. We also study the form of the possible CP violating
currents.

The paper is organized as follows. In section 2 we review the model to
fix the notation used along the paper. In section 3 we derive the
general Ward identities from the NJL generating functional, and in
particular, the form of the chiral anomaly which appears in the model.
In section 4 the regularization in Minkowski space is described.
In section 5 we describe the counterterms which allow to reproduce the
QCD anomaly as well as their influence on the effective action and
vector and axial currents. The constructive method used to derive the
counterterms is described in detail in section 6. Section 7 is devoted
to the study of the
particular two flavour case with $\sigma, \pi, \rho, A $ and $\omega$
mesons.
In section 8 we present our numerical results for nucleon observables.
Section 9 deals with the effective low energy abnormal parity but not
anomalous action of vector mesons in the presence of external currents.
Finally, in section 10 we summarize our results and present our
conclusions.

{\bf 2. The NJL Model and Bosonization Revisited }

Our starting point is the Nambu--Jona-Lasinio Lagrangian in Minkwoski
space [11]
$$ \eqalign{ \hskip1cm
{\cal L}_{\rm NJL}=
\bar{q} (i\slashchar\partial - \hat{M}_0 )q & +
{G_S \over 2}\sum_{a=0}^{N_f^2-1} \left( (\bar{q}\lambda_a q)^2
                        +(\bar{q}\lambda_a i \gamma_5 q)^2 \right) \cr &
-{G_V \over 2}\sum_{a=0}^{N_f^2-1} \left(
(\bar{q}\lambda_a \gamma_\mu q)^2
  +  (\bar{q}\lambda_a  \gamma_\mu \gamma_5 q )^2 \right)
\cr }
\eqno(2.1)$$
where $q=(u,d,s, \dots )$ represents a quark spinor with $N_c $ colours
and $N_f$
flavours. The $\lambda$'s represent the Gell-Mann matrices of the
$U(N_f)$ group (see Appendix A) and  $\hat M_0= {\rm diag} (m_u, m_d,
m_s, ...) $ stands for the current quark mass matrix.
In the limiting case of vanishing current quark masses the NJL-action
is invariant under the global $U(N_f)_R \otimes U(N_f)_L $ group of
transformations (see e.g. Appendix A). The corresponding vector and
axial currents are given by
$$ J_{\mu a}^V (x) = {1\over 2} \bar q(x) \gamma_\mu \lambda_a q(x);
\qquad  J_{\mu a}^A (x) = {1\over 2} \bar q(x) \gamma_\mu \gamma_5
   \lambda_a q(x);
\eqno(2.2) $$
respectively. We will not consider the effects of a $U_A (1)$
breaking term as done in [35] since they are not relevant for
what follows. In order not to overload the paper with notation we will
always work in Minkowski space and will never specify the Wick
rotation explicitly. In fact well defined results can be obtained by
using the customary replacement $\hat M_0 \to \hat M_0 - i\epsilon $.
Nevertheless all our results can be equivalently obtained from a
Euclidean analytical continuation following the
conventions of appendix B.

The vacuum to vacuum transition amplitude in the presence of external
bosonic $(s,p,v,a)$ and fermionic $(\eta, \bar\eta) $ fields of the NJL
Lagrangian can be written as a path integral as
$$ \eqalign{ Z[s,p,v,a,\eta,\bar\eta] &=
\langle 0 | {\rm T} \exp \Bigl\{ i \int d^4 x
\[\bar q \({\slashchar v}+{\slashchar a} \gamma_5 -(s+i\gamma_5 p)\)q
+ \bar \eta q + \bar q \eta \] \Bigr\} |0 \rangle \cr &
= \int D\bar{q} Dq \exp \Bigl\{i\int d^4 x \[ \L_{\rm NJL} +
\bar q \({\slashchar v}+{\slashchar a} \gamma_5 -(s+i\gamma_5 p)\)q
+ \bar \eta q + \bar q \eta \] \Bigr\} \cr}
\eqno(2.3)$$
Following the standard procedure [36] it is convenient to introduce
auxiliary bosonic fields so that one gets the equivalent generating
functional
$$
Z[s,p,v,a,\eta,\bar\eta] = \int D\bar q D q DS DP DV DA \,{\rm
exp}\Bigl\{i\int d^4 x
{\cal L}_{\rm sm}(x)  \Bigr\}
\eqno(2.4)$$
where the semi-bosonized Lagrangian reads
$$ \L_{\rm sm} = \L_{\rm int} + \L_{\rm ext}^F + \L_{\rm ext}^B + \L_m
+\L_{M_0}
\eqno(2.5)$$
and
$$\eqalign{
& \L_{\rm int} = \bar q \( {\slashchar{ V}} + {\slashchar{ A}}
\gamma_5 - ( S+ i \gamma_5 {     P}) \) q = -\bar q M_{\rm int} q \cr
& \L_{\rm ext}^B =\bar q \( i \slashchar\partial + {\slashchar{v }} +
                  {\slashchar{a}} \gamma_5 - ( s+ i \gamma_5 p ) \) q =
\bar q i\D_{\rm ext} q \cr
& \L_{\rm ext}^F = \bar{\eta} q + \bar{q} \eta \cr
& \L_m = -{1\over 4G_S} \tr(S^2 + P^2) + {1\over 4G_V} \tr( V_\mu V^\mu
+ A_\mu A^\mu )  \cr
& \L_{M_0} = - \bar q \hat{M}_0 q
\cr}
\eqno(2.6)$$
Here $(S,P,V,A)$ are dynamical internal bosonic fields, whereas $(s,p,v,a)$ and
$(\eta,\bar\eta)$ represent external bosonic and fermionic fields
respectively.  The bosonic fields are all of them expanded in terms of
the $\lambda$ flavour matrices (See Appendix A). Notice also that for
the path integral in the bosonic fields to be well defined in
Minkowski space we must
use the prescription $ {1\over G_S} \to {1\over G_S} -i \epsilon $ and
$ {1\over G_V} \to {1\over G_V} - i\epsilon $.
Finally, if fermions are formally integrated out one obtains
$$ Z[s,p,v,a,\eta,\bar\eta ] = \int DM_{\rm int} \Det (i\D )
\exp\Bigl\{ -i \langle \bar \eta, (i\D)^{-1} \eta \rangle \Bigr\}
\exp\Bigl\{ i\int d^4 x \L_m \Bigr\}
\eqno(2.7)$$
where the bosonic integration measure $DM_{\rm int} = DSDPDVDA$ and the
usual notation
$$ \langle \bar \eta, (i\D)^{-1} \eta \rangle = \int \int d^4 x d^4 x'
\bar\eta   (x)  \langle  x| (i\D+i\epsilon)^{-1}| x' \rangle \eta(x')
\eqno(2.8)$$
have been introduced.
Concerning the previous manipulations and anticipating the forthcoming
sections, a few remarks should be made.
The path integral above is a highly singular object and defines a whole
family of effective physical theories depending on the particular
prescriptions employed to give it a meaning. This includes in
particular the
regularization procedure and the symmetries respected by it.
In the absence of a real derivation from QCD of the NJL model,
one can only hope that such a freedom can be used to reproduce as many
known features of the underlying fundamental theory as possible. This
set of possible choices is, however, not completely arbitrary. It only
reflects the ambiguity arising in the regularization of a certain
ultraviolet divergent Feynman diagram, namely the fact that in a loop
graph the divergent piece is a polynomial in the external incoming
momenta and also that only diagrams with less than five legs in four
space-time dimensions are divergent. The coefficients of such a
polynomial depend on the regularization. In practice, this means that
depending on the regularization one can always subtract local and
polynomial counterterms in the fields to the action. In this paper we
are interested in the NJL model as an effective theory of QCD and in the
possible counterterms that allow to reproduce the QCD chiral anomaly for
a given regularization prescription.

It should be kept in mind that, strictly speaking, the bosonization
procedure is not obligatory. Nevertheless it will prove highly
convenient in what follows since it reorders the diagrammatic expansion
in a convenient way, so that $N_c$ counting rules become self-evident.
More important, it is the
only known device to describe in practice baryons as chiral solitons.
Nevertheless, the bosonization may appear to impose some conditions on
the
theory, since it treats classes of infinite graphs on the same footing.
This point will be discussed later in more detail.

{\bf 3. Ward Identities }

For the moment we will ignore the regularization and proceed formally.
The resulting expressions will only acquire a precise meaning when the
mathematical objects involved are provided with a suitable
regularization. For simplicity we will consider the chiral limit, i.e.
we set $\hat{M}_0 =0$, since their influence is expected to be small in
the present context.

To obtain the form of Ward identities we decompose first the Dirac
operator into an internal plus external field contribution both
transforming homogeneously under local chiral transformations, i.e. we
define
$$ i\D = i\D_{\rm ext} - M_{\rm int}
\eqno(3.1)$$
and
$$\S = s + S ; \qquad
  \P = p + P ; \qquad
  \V = v + V ; \qquad
  \A = a + A
\eqno(3.2)$$
A local chiral rotation of the external fields
induces the change $ \D_{\rm ext} \to \D_{\rm ext}^g $
(see Appendix A for details and conventions). Let us
call $ Z^g [s,p,v,a,\eta,\bar\eta] =
Z [ s^g , p^g , v^g , a^g , \eta^g, \bar{\eta}^g ] $
the generating functional so transformed. We make a change of variables
in the dynamical bosonic
fields $ (S,P,V,A) \to (S^g , P^g , V^g , A^g ) $ which in the
infinitesimal case reads
$$\eqalign{
& \delta V^\mu = i[ \epsilon_V , V^\mu ] + i [ \epsilon_A , A^\mu ] \cr
& \delta A^\mu = i [ \epsilon_V , A^\mu ] + i[ \epsilon_A, V^\mu ] \cr
& \delta S = i [ \epsilon_V , S ] + \{ \epsilon_A , P \} \cr
& \delta P = i [ \epsilon_V , P ] - \{ \epsilon_A , S \} \cr }
\eqno(3.3)$$
Notice that the transformation is local but homogeneous for all the
internal fields, i.e. no derivative terms appear. This is required if
$\V,\A$
and $v,a$ have to transform non-homogeneously. The bosonic measure is
invariant under these local homogeneous chiral transformations
$ DS DP DV DA = DS^g DP^g DV^g DA^g $ and also the bosonic mass terms
$\L_m^g = \L_{m} $. This property allows to freely choose whether the
bosonic fields are transformed or not, i.e. whether we take
$i\D^g = i\D^g_{\rm ext} - M_{\rm int}^g $ or
$i\D^g = i\D^g_{\rm ext} - M_{\rm int} $  respectively.
We have
$$ Z^g [ s , p, v, a,\eta,\bar{\eta}]
= \int D\bar q D q D M_{\rm int} \exp\Bigl\{ i\int d^4 x \[
\bar q i\D^g q + {\cal L}_m + \bar{q} \eta^g +
\bar{\eta}^g q \] \Bigr\}
\eqno(3.4)$$
Integrating out the fermions we get
$$ Z^g [ s , p, v, a,\eta,\bar{\eta}]
=   \int DM_{\rm int} \Det (i\D^g )
\exp\Bigl\{ -i \langle \bar\eta, (i\D)^{-1} \eta \rangle \Bigr\}
\exp\Bigl\{ i\int d^4 x {\cal L}_m \Bigr\}
\eqno(3.5)$$
Note that the term  $\langle \bar\eta, (i\D)^{-1} \eta \rangle $ given
by eq.(2.8) is invariant under the local chiral transformations
specified above. For infinitesimal transformations we have
\footnote{ ${}^* $ } { The total trace Sp is made out of the
space-time trace, the colour
trace, the Dirac spinor trace ${\rm tr}_\gamma $ and the flavour trace
tr. }
$$ \delta Z[s,p,v,a,\eta,\bar\eta ] = \int DM_{\rm int}
\delta \( {\rm Sp}\, \log(i\D) \) \Det (i\D )
\exp\Bigl\{ -i \langle \bar \eta, (i\D)^{-1} \eta \rangle \Bigr\}
\exp\Bigl\{ i\int d^4 x \L_m \Bigr\}
\eqno(3.6)$$
Decomposing the
variation into its vector and axial parts as follows
$$\delta \( {\rm Sp}\, \log(i\D) \) = 2i \sum_{a=0}^{N_F^2 -1} \int
d^4 x [ \epsilon_V^a (x) \A_V^a (x) +
\epsilon_A^a (x) \A_A^a (x) ]
\eqno(3.7)$$
and compare with the corresponding variation before integrating
out the quarks we get the following identities
$$ \eqalign{ \partial^\mu J^V_{\mu a} & = -\A_V^a (x) + \cr  &
+ f_{abc} \[ J^V_{\mu b} v^\mu_c + J_{\mu b}^A a^\mu_c \]
-{1\over 2} f_{abc} \[  \bar q \lambda_b q s_c +
\bar q i\lambda_b \gamma_5 q  p_c \]   + {i\over 2}\[ \bar q \lambda_a
\eta  - \bar \eta \lambda_a q \]  \cr
\partial^\mu J^A_{\mu a} & = -\A_A^a (x) + \cr &
+ f_{abc} \[ J^V_{\mu b} a^\mu_c + J_{\mu b}^A v^\mu_c \]
+{1\over 2} d_{abc} \[  \bar q \lambda_b q s_c -
\bar q i\lambda_b \gamma_5 q  p_c \]
- {1\over 2}\[ \bar q i\lambda_a
\gamma_5 \eta  + \bar \eta i\lambda_a \gamma_5 q \] \cr }
\eqno(3.8)$$
This identity is valid under the path
integral weighted with the full semi-bosonized
action given by eq. (2.5). Moreover, it is clear that
since we always obtain the Ward identities by
functional differentiation of an effective action, the Wess-Zumino
consistency conditions [3] are automatically satisfied.

{\bf 4. Effective Action and Regularization}

The determinant of the Dirac operator is an ultraviolet divergent
object. Hence we have to introduce some regularization. In addition,
since the model is non-renormalizable the corresponding cut-off has to
remain finite, at least for the divergent pieces. This is a clear
theoretical uncertainty in the model which has very often been
neglected besides few exceptions. Nevertheless, some constraints can be
imposed on the basis of symmetries.
For the benefit of the reader we
will make  some digression here about the choice of regularization
prescription at the expense of overlapping with
previous works (see specially ref.[29] ).

At a formal level the
determinant of the Dirac operator depends on the sum of the internal
plus external fields. This circumstance is ultimately the reason for the
realization of vector meson dominance through Current-Field Identities
in the model. Naively, one might expect that  after regularization the
result depends on the sum too. This, however, does not necessarily
have to be the case [32].
For instance, one
may first formally expand the effective action in powers of the fields
and apply a regularization prescription afterwards. Proceeding in this
way one has the freedom to regularize each vertex separately in a way
that additivity is not fulfilled.  One can add to the
Lagrangian the most general local
counterterm of at most mass dimension four which does not depend on the
sum of internal plus external fields as follows
$$
\log \Det ( i\D ) := \log \overline{\Det} ( i\D ) + i \Delta \Gamma
[v,a;V,A]
\eqno(4.1)$$
where the bar stands for a vector additive regularized fermion
determinant and $\Delta \Gamma$ are the counterterms. Such
a decomposition is convenient within the chiral soliton approach to
baryonic structure. In the NJL model, the regularized determinant
represents the contribution of the polarized Dirac sea to baryonic
observables and is usually evaluated as a regularized sum of
eigenvalues, thus conserving additivity. This corresponds to
the $\log \overline{\Det} ( i\D ) $ piece. In the next section we
will discuss the constraints on the counterterms $\Delta \Gamma$.
In the remainder of this section we give our precise definition of
$\log \overline{\Det} ( i\D ) $.

The additive contribution to the effective action can be separated into
a  $\gamma_5$-odd and $\gamma_5$-even part. It is convenient to
introduce the operator
$$ \D_5 [ \S, \P, \V, \A ] = \gamma_5 \D [ \S,-\P,\V,-\A] \gamma_5 =
-\D [- \S,\P,\V,-\A]
\eqno(4.2)$$
In fact, $\D_5 $ corresponds to rotate $\D$ to Euclidean space, take
the hermitean conjugate and rotate back to Minkowski space (see
Appendix B). This definition allows to separate the action into a
$\gamma_5$-even part (normal pseudoparity) and a $\gamma_5$-odd part
(abnormal pseudoparity). The former can be regularized in a chiral gauge
invariant manner by means of the Pauli-Villars scheme [37]
$$ \eqalign{ \log \overline{\Det}|_{\rm even} &= {1\over 2} \[
{\rm Sp log}(i\D+i\epsilon) +
{\rm Sp log}(i\D_5 +i\epsilon) \]
\cr &= {1 \over 4} \[ {\rm Sp log}(\D \D_5 +i\epsilon ) +
{\rm Sp log} (\D_5\D +i\epsilon)\] \cr &
\to {1 \over 4} {\rm Sp}
\sum c_i
\[ \log(\D \D_5 + \Lambda_i^2 +i\epsilon) + \log (\D_5 \D +
\Lambda_i^2 +i\epsilon)\]  \cr }
\eqno(4.3)$$
where the Pauli-Villars regulators fulfill $c_0=1$, $\Lambda_0 =0$ and
$\sum_i c_i =0$, $\sum_i c_i \Lambda_i^2 =0 $.
For the $\gamma_5$-odd part we formally have
$$\eqalign{
 \log \overline{\Det}|_{\rm odd} &
 = {1\over 2} \[ {\rm Sp log}(i\D+i\epsilon) - {\rm Sp
log}(i\D_5+i\epsilon) \] \cr
 &= {1 \over 4} \[ {\rm Sp log}(\D^2-i\epsilon )
-{\rm Sp log} (\D_5^2-i\epsilon )\]\cr}
\eqno(4.4)$$
The main difference to the $\gamma_5$-even part is that the sum of the
eigenvalues implied in eq. (4.4) is conditionally convergent without the
need of an explicit cut-off. The result is however unambiguous if one
further imposes reproducing the additive anomaly $G_A[\V,\A]$ (see
eq. (5.4)). This is a consequence of the
absence of vector-gauge invariant fourth order terms of abnormal
pseudoparity. A practical formal expression suitable for a heat-kernel
expansion is given by
$$
 \log \overline{\Det}|_{\rm odd} =
 -{1\over 4} \int_0^\infty {d \alpha\over \alpha}
 {\rm Sp} \[ e^{-i\alpha( \D^2 - i\epsilon)}
      - e^{-i\alpha( \D_5^2 - i\epsilon)} \]
\eqno(4.5)$$
For this expression to be well defined one must take the limit
$\epsilon \to 0^+ $ at the end of the calculation.

Finally, it should be mentioned that our separation into odd and
even parts implies in itself a regularization procedure. Instead one
might consider a vector additive regularization applied directly to the
determinant of the Dirac operator, considering the operator $\D$ only.
The corresponding action so regularized would differ in general from
ours by vector gauge invariant, additive and chirally breaking
polynomial counterterms whose coefficients may depend on the
regularization. This feature is due to the fact that regularization
and separation into odd and even parts are not commuting operations and
reflects once more the arbitrariness in the definition of the fermion
determinant. Thus, if such a kind of regularization procedure were used
these additional counterterms should have to be subtracted.

{\bf 5. Chiral Anomaly, Counterterms and Currents in Minkowski space}

Under a local chiral rotation the variation of the Dirac determinant
can be separated into two contributions. Due to the transformation
properties
$$\eqalign{ &
\delta( \D \D_5 ) =i [  \epsilon_V-\epsilon_A \gamma_5, \D \D_5 ] \,;
\qquad
   \delta( \D_5 \D ) =i [ \epsilon_V+ \epsilon_A \gamma_5, \D_5 \D] \cr
& \delta(\D^2 ) = +i [\epsilon_V , \D^2 ]
- i \{ \{ \epsilon_A \gamma_5 , \D \} , \D \}\,;
\qquad
  \delta(\D_5 ^2 ) = +i [\epsilon_V , \D_5^2 ]
+ i \{ \{ \epsilon_A \gamma_5 , \D_5 \} , \D_5 \}  \cr }
\eqno(5.1)$$
and using the trace cyclic property, valid under regularization, we get
$$
\delta \log \overline{\Det}|_{\rm even} = 0
\eqno(5.2)$$
for the even piece whereas for the odd piece we obtain
$$
\delta \log \overline{\Det}|_{\rm odd} =-i \lim_{\eta\to 0^+}
{\rm Sp } \[ \epsilon_A \gamma_5
\( e^{-i \eta \D^2 } + e^{ -i\eta \D_5^2 }\)\]
\eqno(5.3)$$
Straightforward calculation using the usual heat kernel method
[38] yields
the following result for the infinitesimal change of the regularized
determinant under local chiral transformations
$$
\delta \log \overline{\Det} (i\D) = + i \int d^4 x \tr [ \epsilon_A
(x) \A_A (x) ] = i G_A [ \V , \A ]
\eqno(5.4)$$
where
$$\eqalign{
\A_A (x) = {N_c \over 4\pi^2 } & \epsilon_{\mu\nu\alpha\beta}\Bigl\{
{1\over 4} \F^{\mu\nu} \F^{\alpha\beta} +
{1\over 3} \A^\mu \A^\nu \A^\alpha \A^\beta + \cr & +
{i\over 6} \{ \F^{\mu\nu} , \A^\alpha \A^\beta \} +
{2i\over 3} \A^\mu \F^{\nu\alpha} \A^\beta +
{1\over 3} [ {\cal D}^\mu , \A^\nu ] [ {\cal D}^\alpha,\A^\beta ]\Bigr\}
\cr }
\eqno(5.5)$$
with
$$ {\cal F}_{\mu\nu}= \partial_\mu \V_\nu - \partial_\nu \V_\mu -
i[\V_\mu , \V_\nu] =i [ {\cal D}_\mu , {\cal D}_\mu ]
\eqno(5.6)$$
Notice that the fields appearing in the expression of the anomaly are
the sum of internal plus external fields. The QCD anomaly in its
Bardeen form depends on the external fields only and corresponds to
put in eq. (5.5) $\A= a $ and $\V=v$ or correspondingly $V=0$ and $A=0$.
Hence the NJL model reproduces the proper anomaly
\footnote { ${}^*$ } {
There is a sign difference with the work of Bardeen [1] due to the
different convention for the Levi-Civita tensor $ \epsilon_{0123} ({\rm
Bardeen} ) = + 1 $ whereas we have $ \epsilon^{0123} = - \epsilon_{0123}
= + 1 $ .}
if $G_V=0$ and $\Delta\Gamma =0$ defined by eq. (4.1). This is the
solution found in our previous
paper [23]. However, this is not the only solution. To keep the line
of reasoning straight we anticipate the result to be derived in the
next section.  Another solution (unique up to CP
violating terms) is given by $G_V \neq 0 $ and
$$\eqalign{&
\Delta \Gamma = -{iN_c \over 24\pi^2} \int \tr \(
6ia \{ F,V \} + 3iF [A,V] + 4a^3 V + a^2 [A,V] + 2a \{ A^2 , V\}
+ 4 a V a A \cr & +
4 a V^3 + 2i Da [a,A] + i DA [a,A] + 3i DV[a,V] +
2iDV [A,V] - VA^3 - 3 V^3 A  \) \cr }
\eqno(5.7)$$
where for convenience we have used notation of differential forms with
the following 1-forms
$$ V=V_\mu dx^\mu ; \qquad A=A_\mu dx^\mu ; \qquad
   v=v_\mu dx^\mu ; \qquad a=a_\mu dx^\mu
\eqno(5.8)$$
and 2-forms
$$ F={1\over 2} F_{\mu\nu} dx^\mu dx^\nu ; \qquad
   DV={1\over 2} (DV)_{\mu\nu} dx^\mu dx^\nu ; \qquad {\rm etc.}
\eqno(5.9)$$
and
$$\eqalign{
& F_{\mu\nu} =\partial_\mu v_\nu -\partial_\nu v_\mu -i[v_\mu,v_\nu]=
i[ D_\mu , D_\nu ] \cr
&(DA)_{\mu\nu} = [ D_\mu , A_\nu ] - [ D_\nu , A_\mu ]\cr
&(DV)_{\mu\nu} = [ D_\mu , V_\nu ] - [ D_\nu , V_\mu ]\cr
&(Da)_{\mu\nu} = [ D_\mu , a_\nu ] - [ D_\nu , a_\mu ]\cr }
\eqno(5.10)$$
and $ dx^\mu dx^\nu dx^\alpha dx^\beta = d^4 x
\epsilon^{\mu\nu\alpha\beta} $.  These terms satisfy that
$$\eqalign{ \delta \( -i \log\overline{\Det}(i\D) & + \Delta \Gamma
\)  = G_A [v,a] \cr & =
{N_c \over 4\pi^2 } \int \tr \Bigl\{
\epsilon_A \[ F^2 + {1\over 3} a^4  + {i\over 3} \{ F , a^2 \} +
{4i\over 3} a F a +
{1\over 3}  (D a)^2  ] \] \Bigr\} \cr }
\eqno(5.11)$$
where the fermionic contribution to the effective action is regularized
in a vector gauge invariant manner. We point out that the former
eq. (5.11) corresponds to eq. (5.5) in the particular case $V=0$ and
$A=0$. Notice also that the counterterms
\underbar{ do not} depend on the additive combination $ \V = v + V $ and
$\A= a + A $. Hence we will have corrections to the usual Current-Field
Identities. Furthermore, the counterterms are also written in a
manifestly vector gauge invariant fashion.

The counterterms can be classified according to the number of
external fields. For our purposes only the zeroth order  (modification
of the action $\Delta \Gamma_0$) and first order (modification of the
currents $\Delta J_V $ and $\Delta J_A$) will be needed. They are
$$\eqalign{ \Delta \Gamma_0 =-{i N_c \over 24\pi^2 }&
\int \tr \[  2 i dV [ A , V] -  V A^3 - 3 V^3 A  \] \cr
 \Delta \Gamma_1 = -{iN_c \over 24 \pi^2} & \int
 \tr \[ v \( 3i \{ dA , V \}
-3i \{ dV,A\} + 4 VAV - 2\{ A,V^2 \} \) + \cr & + a \( 3i \{ dV
, V \} +i \{ dA,A \} + 4 V^3 + 2\{ A^2 ,V \} \) \] = \cr = &
\int \tr\( v \Delta J_V + a \Delta J_A \) \cr  }
\eqno(5.12)$$
The first order term can be also expressed in another form if use is
made of the self-consistent equations of motion.
The total action in Minkowski space can be written as
$$
W =  \overline{W} [ s + S , p + P , v + V , a+ A ] + \Delta \Gamma [v,a;
V,A]
+ W_m [ S,P,V,A]- \langle \bar \eta, (i\D)^{-1} \eta \rangle
\eqno(5.13)$$
where we have emphasized the fact that in our particular regularization
the fermion determinant $ \overline{W} = -i\log\overline{\Det}(i\D) $
depends on the additive combinations $ s + S $, etc. The term in the
external fermionic fields depends also on these additive combinations.

The former expressions can be brought into a more appealing form by
considering a saddle point approximation of the action (which in this
model becomes exact in the large $N_c$ limit). In effect, if we minimize
the total action with respect to the dynamical fields $V$ and
$A$ (the minimization with respect to $S$ and $P$ is not relevant in
what follows)
in the presence of external fields we find that under the path integral
$$
{\delta  \overline{W}  \over \delta V} +
{\delta \Delta \Gamma \over \delta V} + {1\over 2G_V} V -
{\delta \over \delta V} \langle \bar \eta, (i\D)^{-1} \eta \rangle = 0
\eqno(5.14)$$
and similarly for $A$.
On the other hand, the total currents are
$$\eqalign{ J_V & =
{\delta  \overline{W}  \over \delta v} +  {\delta \Delta \Gamma \over
\delta v}
- {\delta \over \delta v} \langle \bar \eta, (i\D)^{-1} \eta \rangle
\cr & =
{\delta  \overline{W}  \over \delta V} + {\delta \Delta \Gamma \over \delta v}-
{\delta \over \delta V}\langle \bar \eta, (i\D)^{-1} \eta\rangle \cr &=
-{1\over 2G_V} V + \({\delta \over \delta v}-
  {\delta \over \delta V} \) \Delta \Gamma \cr }
\eqno(5.15)$$
where in the last step we have used the equations of motion (5.14). For
definiteness we will refer to them as the self-consistent currents.
Notice that through the equations of motion the dynamical fields
acquire a dependence on the external bosonic and fermionic fields.
One can prove that the implicit dependence on  the external fermionic
fields can be neglected in the
limit $N_c \to \infty$ as well as the bosonic integration. Thus we get
in this limit
$$\eqalign{ &  \langle 0|T \[ J_V (x) \exp\Bigl\{ i\int d^4 x \L_{\rm
ext}  \Bigr\}\]|0 \rangle
\to \cr &
\to \( -{1\over 2G_V} V + \[{\delta \over \delta v}-
            {\delta \over \delta V} \]\Delta\Gamma \)
\langle 0|T \exp\Bigl\{ i\int d^4 x \L_{\rm
ext}  \Bigr\}\]|0 \rangle \cr }
\eqno(5.16)$$
If we set the external bosonic fields equal to zero then we get
that the total currents are given by
$$\eqalign{ J^V_\mu  = & -{1\over 2G_V} V_\mu + \cr &
- {iN_c \over 24\pi^2}\epsilon_{\mu\nu\alpha\beta}
\[ i \{ \partial^\nu A^\alpha , V^\beta \} +
i\{\partial^\nu V^\alpha ,A^\beta \} + A^\nu A^\alpha A^\beta +
\{ V^\nu V^\alpha , A^\beta \} + V^\nu A^\alpha V^\beta  \] \cr
           J^A_\mu  = & -{1\over 2G_V} A_\mu + \cr &
- {iN_c \over 24\pi^2}\epsilon_{\mu\nu\alpha\beta}
\[i\{ \partial^\nu  V^\alpha ,V^\beta \}
+i\{ \partial^\nu A^\alpha ,A^\beta \} + \{ V^\nu ,A^\alpha A^\beta \}
+ V^\nu V^\alpha V^\beta + A^\nu V^\alpha A^\beta \]
\cr}
\eqno(5.17)$$
These equations represent the leading $N_c$ modifications to
the usual Current-Field Identities [22] and as we see they are valid
in the presence of external quark fields (see eq. (5.16)). In
particular, they can be used to
evaluate baryon matrix elements or form factors. It is interesting to
notice that in terms of right and left field representation (see
Appendix A) these equations can be rewritten as
$$\eqalign{
& J_\mu^R = -{1\over 2G_V } V_\mu^R + {N_c \over 24\pi^2}
\epsilon_{\mu\nu\alpha\beta}\[  {1\over 2} \{ V_R^{\nu \alpha},
V_R^\beta \} + i V_R^\nu V_R^\alpha V_R^\beta \] ; \cr
& J_\mu^L = -{1\over 2G_V } V_\mu^L + {N_c \over 24\pi^2}
\epsilon_{\mu\nu\alpha\beta}\[  {1\over 2} \{ V_L^{\nu \alpha},
V_L^\beta \} + i V_L^\nu V_L^\alpha V_L^\beta \] ; \cr }
\eqno(5.18)$$
where
$$
V_R^{\mu\nu}=\partial^\mu V_R^\nu -\partial^\mu V_R^\nu -
i[ V_R^\mu , V_R^\nu ] ;  \qquad
V_L^{\mu\nu}=\partial^\mu V_L^\nu -\partial^\mu V_L^\nu -
i[ V_L^\mu , V_L^\nu ]
\eqno(5.19)$$
Let us remind that in writing the former expressions explicit use of
the equations of motion has been made. It is interesting to note that
the correction to the current field identities coincide with
the difference between the covariant and consistent currents related
to the covariant and consistent forms of the anomaly respectively
[33]. Thus, these currents do not posses an internal anomaly.

{\bf 6. Counterterms from the Gauged Wess-Zumino action}

In this section we describe the constructive method that we have used to
compute the counterterms $\Delta \Gamma $ given by eq. (5.7).  We remind
that they have to be summed to the additive vector gauge invariant
fermionic action $-i\log\overline{\Det} (i\D) $ so that the total sum
reproduces the QCD anomaly. To do so we will proceed by considering the
gauged Wess-Zumino term. This represents no limitation since the anomaly
is saturated by it. In addition, it should be kept in mind that the
final expression for $\Delta \Gamma $ has to be a polynomial in the
fields.

The left-right gauged Wess-Zumino term
as a functional of the additive field combinations $\A_R$ and $\A_L $
reads (see  e.g. [34] for an explicit construction)
$$\eqalign{
\Gamma_{\rm WZ}^{\rm RL} [ U, \A_R , \A_L ] &= \Gamma_{\rm WZ} [U] \cr
& + {N_c \over 48\pi^2} \int
\tr\[ \A_R U_R^3 + i\{ \A_R , d\A_R \} U_R + i U^\dagger \A_L U \A_R
U_R^2 \cr  &  +i U^\dagger \A_L U d \A_R U_R + {i\over2} (\A_R U_R )^2
+ \A_R^3 U_R - U^\dagger \A_L U \{ \A_R , d\A_R \} \cr &
- U^\dagger \A_L U \A_R U_R \A_R + iU^\dagger \A_L U \A_R^3
+ {i\over 4} ( U^\dagger \A_L U \A_R )^2 \] -{\rm p.c.} \cr }
\eqno(6.1)$$
where p.c. means interchanging the $R$ and $L$ labels and $\Gamma_{\rm
WZ}[U]
$ represents the topological Wess-Zumino action [4] (see
Appendix D) and $U$ a unitary
flavour field $U^\dagger U = 1$. Here the following
1-forms have been defined
$$ \A_R = v_R + V_R ; \qquad \A_L = v_L + V_L ; \qquad
U_R = U^\dagger dU ; \qquad U_L = U dU^\dagger
\eqno(6.2)$$
Under a chiral transformation the fields transform as follows
$$\eqalign{
& \delta v_R = d \epsilon_R + i[\epsilon_R, v_R] ;\qquad
  \delta v_L = d \epsilon_L + i[\epsilon_L, v_L] \cr
& \delta V_R =              + i[\epsilon_R, V_R] ; \qquad \qquad
 \delta V_L =              + i[\epsilon_L, V_L] \cr
& \delta U = i ( \epsilon_L U - U \epsilon_R ) ; \qquad
 \delta U^\dagger =i(\epsilon_R U^\dagger -U^\dagger \epsilon_R ) \cr }
\eqno(6.3)$$
i.e., the dynamical fields transform homogeneously, whereas the external
fields transform non-homogeneously. The variation gives the right-left
form of the anomaly
$$ \delta \Gamma_{\rm WZ}^{\rm RL}[U,\A_R,\A_L ] = {N_c \over 48\pi^2 }
\int
\tr\[ \( \{ \A_R , d\A_R \} - i \A_R^3 \) d\epsilon_R \] - {\rm p.c.}
\eqno(6.4)$$
To bring this anomaly to a vector gauge invariant form we consider
the polynomial action
$$\eqalign{
& \Gamma_{\rm WZ}^{\rm RL} [ 1, \A_R , \A_L ] = {N_c \over 48\pi^2}
\int \tr\[ - \A_L \{ \A_R , d\A_R \}  + i \A_L \A_R^3
+ {i\over 4} ( \A_L \A_R )^2 \] -{\rm p.c.} \cr }
\eqno(6.5)$$
where $1$ means the unit matrix in flavour space ($U=1$ in (6.1)).
The anomaly (5.4) can be reproduced by the effective action
$$ \Gamma_{\rm WZ}^V [ U, \A_R , \A_L ] =
\Gamma_{\rm WZ}^{\rm RL}[U,\A_R,\A_L ]
-\Gamma_{\rm WZ}^{\rm RL}[1,\A_R,\A_L ] ; \qquad \delta
\Gamma_{\rm WZ}^V [U, \A_R , \A_L ] = G_A [ \V , \A ]
\eqno(6.6)$$
This action coincides with that of ref. [23] and any other action
reproducing the anomaly in terms of additive fields (5.4) differs by
chirally invariant terms from this one. Actually, the former action
(6.6) is the leading contribution in a gradient expansion of the
$\gamma_5$-odd part of $\log \overline{\Det} (i\D)$ provided the
non-linearly transforming field $U$ is taken as the (unique) unitary
part of
the flavour matrices $S+iP$. Hence, the terms of action (6.6) containing
the fields either $U$ or $U^\dagger $ are not polynomial actions due to
the chiral circle condition $U^\dagger U = 1$.  This anomaly,
however, \underbar{does not} coincide with the QCD anomaly [1],
since it contains the dynamical vector and axial fields $V$ and $A$.
As we know the QCD anomaly depends on the external fields $v$ and
$a$ only. To eliminate the internal fields dependence we propose the
most general globally chiral invariant counterterm depending on vector
and axial degrees of freedom in a way that their variation exactly
cancels the dependence on the internal fields in the anomaly.
$$\eqalign{ \Gamma^{\rm RL}_{\rm ct} [v_R, v_L ; V_R , V_L ] & = \int
\tr\Bigl\{ c_1 V_R v_R^3 +
c_2 V_R dv_R v_R + c_3 V_R v_R dv_R + c_4  V^2_R v_R^2 \cr &
+ c_5  ( V_R v_R )^2 + c_6 V_R^2 dv_R
+c_7 dV_R V_R v_R + c_8 V_R^3 v_R  \Bigr\} -{\rm p.c.} \cr }
\eqno(6.7)$$
Under a chiral transformation we have
$$\eqalign{ \delta\Gamma_{\rm ct}^{\rm RL} = \int &
\tr\Bigl\{ \[ (c_1 + ic_2) v_R^2
V_R + (c_1 + ic_3 ) V_R v_R^2 -(c_1 + ic_2 +ic_3 )v_R V_R v_R \cr &
+ c_2 V_R dv_R + c_3 dv_R V_R -(c_4 + i c_6 ) v_R V_R^2 + (c_4 + ic_6
-ic_7) V_R^2 v_R \cr & +
(2c_5 + ic_7) V_R v_R V_R + c_7 dV_R V_R + c_8 V_R^3 \] d\epsilon_R
\Bigr\} - {\rm p.c.} \cr }
\eqno(6.8)$$
The coefficients $c_1 , \dots , c_8$ are to be fixed by imposing that
$$ \delta \( \Gamma_{\rm WZ}^{\rm RL}[U,\A_R , A_L
]-\Gamma_{\rm ct}^{\rm RL}[v_R,v_L;V_R,V_L ] \)
= {N_c \over 48\pi^2}\int
\tr\[ \( \{ v_R , d v_R \} - i  v_R^3 \) d\epsilon_R \] - {\rm p.c.}
\eqno(6.9)$$
This equation fixes all coefficients except one
$$\eqalign{ \Gamma^{\rm RL}_{\rm ct} [v_R,v_L;V_R,V_L] = -{i N_c \over 48\pi^2
}
\int \tr & \[ 3 V_R v_R^3 + 2i V_R \{ dv_R , v_R \}
+ c\, V^2_R ( dv_R - iv_R^2 ) \cr & + i \{ dV_R , V_R \}  v_R
+ {3\over 2} ( V_R v_R )^2 + V_R^3 v_R \] -{\rm p.c.} \cr }
\eqno(6.10)$$
The undetermined coefficient $c$ vanishes if, in addition, we impose CP
invariance. Indeed, under CP we have
$$ {\rm CP}:\, (V_R)_\mu ( t, \vec x ) \to - (V_L)^{\mu, t} ( t, -\vec
x); \qquad (V_L)_\mu ( t, \vec x ) \to - (V_R)^{\mu, t} ( t, -\vec x)
\eqno(6.11)$$
where upperscript $t$ means matrix-transposed. After some integrations
by parts one can see that CP results in changing $ c \to -c $. We
will take $c=0$ for the rest of this section. A brief discussion of CP
violating terms can be found in Appendix E.

The following action reproduces the correct anomaly (5.11)
$$ \Gamma_{\rm WZ}^V [ U,v,a; V,A] =
\Gamma_{\rm WZ}^{\rm RL}[U,\A_R,\A_L ]
-\Gamma_{\rm ct}^{\rm RL}[v_R,v_L; V_R,V_L ] -
\Gamma_{\rm WZ}^{\rm RL} [ 1, v_R , v_L ]
\eqno(6.12) $$
i.e. $\delta \Gamma_{\rm WZ}^V [ U,v,a; V,A] = G_A [v,a] $.
However, in a vector gauge invariant calculation of the fermion
determinant the chirally breaking terms, which depend on the additive
combinations $ \A_R = v_R + V_R $ and $ \A_L = v_L + V_L $, are given by
eq. (6.6). Thus we define
$$ \Delta \Gamma = \Gamma_{\rm WZ}^{\rm RL} [ 1 , \A_R , \A_L ]
- \Gamma_{\rm WZ}^{\rm RL} [ 1 , v_R , v_L ] - \Gamma_{\rm ct}^{\rm RL}
[v_R, v_L ; V_R,V_L]
\eqno(6.13)$$
so that
$$ \Gamma_{\rm WZ}^V [ U,v,a; V,A] =\Gamma_{\rm WZ}^V [U,\A_R,\A_L ]
+ \Delta\Gamma [ v,a; V,A ]
\eqno(6.14)$$
Thus, $\Delta \Gamma [v,a;V,A] $ are the counterterms to be added to the
vector gauge invariant fermion determinant $\overline{\Det}(i\D)$,
determined up to local and polynomial chiral gauge invariant
combinations. These combinations must be formed by using the chirally
covariant objects $ F_R$, $DV_R$ and $V_R$ in all
possible combinations. For completeness we repeat here the argument
already given in ref. [32]. If we insist on parity
conservation the following chiral gauge invariant combinations remain
$$
\tr DV_R V_R^2 , \quad
\tr F_R V_R^2, \quad
\tr DV_R F_R , \quad
\tr (DV_R)^2, \quad
\tr V_R^4, \quad
\tr F_R^2
\eqno(6.15)$$
together with the corresponding left-field combinations.
The first term vanishes identically as can be seen integrating by parts.
The second term is the CP violating term already considered before.
The third term vanishes after integration by parts due to the Bianchi
identity for the field strength tensor $ D F_R = 0 $.
The fourth term can be reduced integrating by parts to the second one.
The fifth also vanishes identically due to the cyclic property of the
trace. Finally, the sixth term is a topological action in the external
fields which are assumed to have zero winding and hence vanishes.
This completes the proof that the counterterms are uniquely given if CP
invariance is invoked.

{\bf 7. Study of the two flavour case}

It is interesting to study the two flavour reduction of the model
which corresponds to a NJL Lagrangian of the form
$$\eqalign{  {\cal L}&=
\bar q  (i\slashchar\partial - \hat M_0 ) q +
{G_1 \over 2} \[ (\bar q q)^2 +(\bar q \vec \tau i\gamma_5 q)^2 \] \cr
& - {G_2 \over 2} \[ (\bar q \vec \tau \gamma_\mu q)^2
              + (\bar q \vec \tau \gamma_\mu \gamma_5 q)^2 \]
- {G_3 \over 2} ( \bar q \gamma_\mu q )^2 \cr }
\eqno(7.1)$$
Notice that, strictly speaking, the case $G_2 \neq G_3 $ is not a
particular case of the model considered in section 2, however
trivial modifications can
be easily implemented to the case of interest. After bosonization we get
$$\eqalign{ {\cal L}& =
\bar q \(i\slashchar\partial - g_\pi (\sigma + i\gamma_5 \vec\tau \cdot
\vec \pi ) + {g_\rho \over 2}  \vec \tau \cdot( \vec{\slashchar\rho} +
\vec{\slashchar A}  \gamma_5 ) + g_\omega \slashchar\omega - \hat M_0 \)
q \cr &
- {1\over 2} \mu^2 (\sigma^2 + \vec \pi^2 ) + {1\over 2} m_\rho^2
( \vec\rho^\mu \cdot \vec\rho_\mu + \vec A^\mu \cdot \vec A_\mu ) +
{1\over
2} m_\omega^2 \omega_\mu \omega^\mu \cr }
\eqno(7.2)$$
with $G_1 =g_\pi^2 / \mu^2 $, $G_2 = g_\rho^2 / (4m_\rho^2) $ and
$G_3= g_\omega^2 /m_\omega^2 $.
Up to the mass term, this Lagrangian is invariant under the $ {\rm
SU}(2)_R \otimes
{\rm SU}(2)_L \otimes {\rm U}_B(1) $ group and has been studied in
detail in refs. [29,30,31]. The corresponding baryon, vector and
axial currents are
$$ J_\mu^B (x) = \bar q(x) \gamma_\mu q(x); \qquad
\vec J_\mu^V (x)={1\over 2}\bar q(x)\gamma_\mu \vec \tau q(x) ; \qquad
\vec J_\mu^A (x)={1\over 2}\bar q(x)\gamma_\mu \gamma_5 \vec \tau q(x)
\eqno(7.3) $$
In the notation of previous sections the reduction of the
Lagrangian corresponds to take the dynamical fields to be
$$
V_\mu = {1\over 2}g_\rho\vec \rho_\mu\cdot\vec \tau + g_\omega
\omega_\mu ;
\qquad A_\mu = {1\over 2} g_\rho \vec A_\mu \cdot \vec \tau
\eqno(7.4)$$
whereas the external fields reduce to
$$
v_\mu = {1\over 2} \vec v_\mu\cdot \vec \tau + v_\mu^0 ; \qquad
a_\mu = {1\over 2} \vec a_\mu \cdot \vec \tau
\eqno(7.5)$$
In this model the Ward identities acquire a simple form in the case of
vanishing $s,p,\eta$ and $\bar\eta$ external fields,
$$ \eqalign{ & \partial^\mu J^B_\mu = 0 \cr &
\partial^\mu \vec J^V_\mu = \vec J^V_\mu
\wedge \vec v^\mu + \vec J_\mu^A \wedge \vec a^\mu \cr
& \partial^\mu \vec J^A_\mu = \vec J_\mu^A \wedge \vec v^\mu +
                              \vec J_\mu^V \wedge \vec a^\mu -
{N_c \over 4\pi^2} \epsilon^{\mu\nu\alpha\beta} \partial_\mu v^0_\nu
\( \partial_\alpha \vec v_\beta + {1\over 2}\vec v_\alpha \wedge
\vec v_\beta - {1\over 2} \vec a_\alpha \wedge \vec a_\beta \) \cr }
\eqno(7.6)$$
The modification of the vector gauge invariant regularized effective
action can then be obtained directly from eqs. (5.12)  giving
$$\eqalign{
\Delta \Gamma_0 = {N_c \over 12\pi^2 } g_\rho^2 g_\omega
\int d^4 x &  \epsilon_{\mu\nu\alpha\beta}\Bigl\{
\partial^\mu \omega^\nu \vec A^\alpha \cdot \vec\rho^\beta +
\partial^\mu \vec\rho^\nu \cdot \vec A^\alpha \omega^\beta \cr &
-{g_\rho \over 8} \omega^\mu [ \vec A^\nu \wedge \vec A^\alpha + 3
\vec\rho^\nu \wedge \vec\rho^\alpha ] \cdot \vec A^\beta \Bigr\} \cr }
\eqno(7.7)$$
In this particular case the total baryon, isospin and axial currents, in
the absence of external fields, become
$$ J_B^\mu = {\delta W \over \delta v_\mu^0 } \Big|_0 \qquad
\vec J_V^\mu = {\delta W \over \delta \vec v_\mu }\Big|_0\qquad
\vec J_A^\mu = {\delta W \over \delta \vec a_\mu }\Big|_0
\eqno(7.8)$$
respectively. Straightforward calculation yields the following results
for the modification of the currents as obtained from eq. (5.12)
$$\eqalign{
\Delta J_\mu^B & = {N_c \over 8\pi^2 } g_\rho^2
\epsilon_{\mu\nu\alpha\beta} \partial^\nu ( \vec A^\alpha \cdot \vec
\rho^\beta) \cr
 \Delta \vec J_\mu^V & = {N_c \over 24\pi^2 } g_\rho g_\omega
 \epsilon_{\mu\nu\alpha\beta} \[ 3 \partial^\nu ( \vec A^\alpha
\omega^\beta
) + 2  g_\rho \omega^\nu ( \vec A^\alpha \wedge \vec \rho^\beta ) \] \cr
\Delta \vec J_\mu^A &= {N_c \over 24\pi^2 } g_\rho g_\omega
\epsilon_{\mu\nu\alpha\beta} \[ 3 \vec \rho^\nu \partial^\alpha
\omega^\beta
+3 \partial^\nu \vec \rho^\alpha \omega^\beta +
g_\rho \omega^\nu ( \vec A^\alpha \wedge \vec A^\beta +
\vec \rho^\alpha \wedge \vec \rho^\beta ) \] \cr }
\eqno(7.9)$$
Notice that the correction to the baryon current is a total divergence,
hence the baryon number normalization is preserved. The total
self-consistent currents, i.e. the total currents
evaluated by means of the equations of motion can be deduced from
eqs. (5.17) and give
\newline
\underbar{ \sl Baryon Current}
$$\eqalign{
 J_\mu^B = -{m_\omega^2 \over g_\omega } \omega_\mu &
+{N_c \over 24\pi^2 } g_\rho^2 \epsilon_{\mu\nu\alpha\beta}\Bigl\{
 \partial^\nu \vec A^\alpha \cdot \vec \rho^\beta +
   \partial^\nu \vec \rho^\alpha \cdot \vec A^\beta +
{1\over 4} g_\rho ( \vec A^\nu \wedge \vec A^\alpha +
3 \vec \rho^\nu \wedge \vec \rho^\alpha )\cdot \vec A^\beta \Bigr\} \cr
}
\eqno(7.10)$$
\underbar{ \sl Vector Current}
$$\eqalign{
&  \vec J_\mu^V = -{m_\rho^2 \over g_\rho} \vec \rho_\mu
+{N_c \over 24\pi^2 } g_\rho g_\omega
\epsilon_{\mu\nu\alpha\beta}\Bigl\{
   \partial^\nu \vec A^\alpha \omega^\beta + \partial^\nu \omega^\alpha
\vec A^\beta + {1\over 2} g_\rho \omega^\nu (\vec \rho^\alpha \wedge
\vec A^\beta )\Bigr\} \cr}
\eqno(7.11)$$
\underbar{ \sl Axial Current}
$$\eqalign{
\vec J_\mu^A = -{m_\rho^2 \over g_\rho} \vec A_\mu &
+{N_c \over 24\pi^2 } g_\rho g_\omega \epsilon_{\mu\nu\alpha\beta}
\Bigl\{ \partial^\nu \omega^\alpha \vec \rho^\beta +
   \partial^\nu \vec \rho^\alpha \omega^\beta +
{1\over 4} g_\rho \omega^\nu ( \vec A^\alpha \wedge \vec A^\beta +
\vec \rho^\alpha \wedge \vec \rho^\beta ) \Bigr\} \cr }
\eqno(7.12)$$
These are the corrected current-field identities in the two flavour case
in
leading order in $N_c$. To conclude this section we notice that if
the fields $\vec \rho_\mu $ and $\vec A_\mu$ vanish, there are no
corrections to any single current.
In contrast, if the field $\omega_\mu$ vanishes there is a correction
to the baryon current not considered in previous works [26,27].

{\bf 8. Numerical Results for Nucleon Observables}

Following the standard approach [11], a baryon can be described in
terms of the corresponding correlation function
$$
\Pi_B (x,x') = \langle 0 | T \Bigl\{ B(x) \bar B(x') \Bigl\} | 0 \rangle
\eqno(8.1)$$
$B(x)$ being a baryonic operator in terms of quark fields. We take
$$
B(x) = {1\over N_c !} \epsilon^{\alpha_1, \dots, \alpha_{N_c}}
\Phi^{a_1,\dots, a_{N_c}} q_{\alpha_1 a_1}(x) \cdots q_{\alpha_{N_c}
a_{N_c}}(x)
\eqno(8.2)$$
where $(\alpha_1, \dots , \alpha_{N_c})$ are colour indices, $(a_1,
\dots, a_{N_c})$ spinor-flavour indices and $\Phi^{a_1,\dots, a_{N_c}}$
the proper completely symmetric spinor-flavour amplitude.
The exact spectral representation of the correlation function is obtained
as usual by inserting the complete set of eigenstates of the NJL hamiltonian
in eq.~(8.1), namely,
$$\eqalign{
\Pi_B (x,x') &= \theta(t-t^\prime)\sum_n\langle 0|B(0)|B_n,\vec{k}\rangle
\langle B_n,\vec{k}|\bar{B}(0)|0\rangle e^{-i(x-x^\prime)k} \cr
&\quad +(-1)^{N_c} \theta(t^\prime-t)
\sum_n\langle 0|\bar{B}(0)|\bar{B}_n,\vec{k}\rangle
\langle \bar{B}_n,\vec{k}|B(0)|0\rangle e^{+i(x-x^\prime)k} \cr}
\eqno(8.3)$$
where $B_n$, ($\bar{B}_n$) are the baryonic (antibaryonic) states with momentum
$\vec{k}$. Further, by chosing the branch $t>t'$ and taking the limit
$t-t' = T \to - i\infty $, the lightest baryon, and at rest, is selected in
the sum,
$$\Pi_B (x,x') = \langle 0|B(0)|B\rangle
\langle B|\bar{B}(0)|0\rangle e^{-iTM_B}
\eqno(8.4)$$
To carry out these steps in the large $N_c$ limit, we first write
the time ordered product (8.1) as a
path integral over fermionic degrees of freedom with weight
${\rm exp}(iS_{\rm NJL})$. The resulting expression,
in turn, can be obtained by appropriate
functional differentiation of the generating functional
$Z[s,p,v,a,\eta,\bar\eta] $ (see eq.(2.3)) with respect to the
external quark fields $\eta(x)$ and $\bar\eta(x)$. After bosonization and
integration of the quarks one gets, using eq.(2.7),
$$
\Pi_B (x,x') =  \Phi^{a_1,\dots, a_{N_c}}
\bar \Phi^{a_1',\dots, a_{N_c}'}
{\int DM_{\rm int} \exp(i\overline{W})\prod_{i=1}^{N_c} iS_{a_i a'_i}(x,x')
\over \int DM_{\rm int} \exp(i\overline{W}) }
\eqno(8.5)$$
where the one particle Green function
$S_{a a'} (x,x') = \langle x| {(i\D+ i\epsilon)}^{-1}_{a a'}|x'\rangle $
has been introduced and all external fields $s,p,v,a$ and $\bar \eta,\eta$
are set to zero. The limit $N_c \to \infty$ drives the functional
integral in the denominator to a saddle point bosonic
configuration which describes
the mean field vacuum. On the other hand, because
there are $N_c$ factors $S_{a,a^\prime}$ in the numerator,
the dominating saddle point configuration will be different
from that of the vacuum and will depend on $x$ and $x^\prime$.
Next, the limit of large evolution time $T$ selects the minimum energy
stationary configuration. For stationary configurations
one can use the spectral representation of the propagator given by
$$ i S_{a a'} ( \vec x , t ; \vec x' , t' ) =
\sum_n (\theta(t-t')\theta(\epsilon_n) -\theta(t'-t)\theta(-\epsilon_n))
\psi_{n a}(\vec x) \bar \psi_{na'}(\vec x')e^{-i \epsilon_n (t-t')}
\eqno(8.6)$$
with $\psi_{n a}(\vec x)$ and $\epsilon_n$ the eigenfunctions and
eigenvalues of the single particle
Dirac Hamiltonian $H$ defined by $i\D = \gamma_0 (i
\partial_t -H)$ (evaluated at the stationary bosonic configuration).
Choosing the $t>t'$ branch
to create a baryon (instead of an antibaryon) and taking the limit
$t-t' = T \to - i\infty $ we get
$$ \Pi_B(x,x') = \Psi_B (\vec x)\bar \Psi_B (\vec x') e^{-iT M_B}
\eqno(8.7)$$
where the baryon wave function
$$ \Psi_B (\vec x) = \Phi^{a_1, \dots, a_{N_c}}
\psi^{\val}_{a_1} (\vec x) \cdots \psi^{\val}_{a_{N_c}} (\vec x)
\eqno(8.8)$$
has been introduced. Here,
the valence level $\epsilon_{\rm val}$ is the lowest positive
eigenvalue of the Dirac Hamiltonian. This level is selected in eq.~(8.6)
by the large Euclidean time limit.
Eq.~(8.7) shows that spatial translational
invariance is spontaneously broken by the mean field approximation,
corresponding to the formation of a solitonic configuration in the presence
of valence quarks.

The baryon mass (in the large $N_c$ limit) comes from
the two contributions in the numerator and denominator,
$M_B = E_B - E_{\rm vac}$, which are given by
$$\eqalign{
E_B &=  \overline{E}_{\rm sea}^{\rm sol}
+ \Delta E^{\rm sol} + E_m^{\rm sol} + N_c \epsilon^{\val} \cr
E_{\rm vac} &=  \overline{E}_{\rm sea}^{\rm vac} + \Delta E^{\rm vac}
+ E_m^{\rm vac} \cr}
\eqno(8.9)$$
where the superscripts ``vac'' and ``sol'' stands for the two different
stationary saddle point configurations.
The valence term (last term) in $E_B$, not present in $E_{\rm vac}$,
comes from the propagators $S_{a,a^\prime}$.
The other terms come from the action
in eq.~(5.13). The sea regularized
and the pure mesonic contribution are given by
$$\eqalign{
\overline{E}_{\rm sea}+ \Delta E  &= {i\over T} \log \Det (i\D)
= {i\over T} \[\log\overline{\Det}(i\D) + i\Delta \Gamma_0 \] \cr
E_m &= -{1\over T} W_m \cr }
\eqno(8.10)$$
respectively. At this point it is clear that any expression for the
total energy involving the Dirac eigenvalues will necessarily depend
on the additive combinations.
Similarly to the energy, one body observables, like e.g. mean squared radius,
axial coupling constant, etc. admit (at leading order in $ N_c$)
a natural decomposition in terms of
a valence part, a pure additively regularized part and the modifications
induced by the counterterms. This fact can be established by repeating similar
steps as for the baryon mass but with non vanishing external bosonic
fields $s,p,v$ and $a$.
As an example we give such a decomposition
for the baryon axial coupling constant
$$ g_A^N = g_A^{\rm val} + \overline{g}_A^{\rm sea} + \Delta g_A^N
\eqno(8.11)$$
i.e. in the absence of counterterms, $\Delta g_A^N$ would vanish.

For stationary fields the Dirac operator can be written as
$$ i\D = \gamma_0 ( i {\partial\over \partial t} - H )
\eqno(8.12)$$
with the one particle Dirac Hamiltonian
$$ H=-i{\alpha\cdot\nabla} - ( V_0 + A_0 \gamma_5 ) +
\alpha\cdot ( V + A \gamma_5 ) + \beta ( S + i\gamma_5 P)
\eqno(8.13)$$
In the case of two flavours
with hedgehog symmetry we have ( $\hat x = \vec x  /
r $ )
$$\eqalign{
& S = g_\pi \sigma (r) ; \qquad P = g_\pi \tau \cdot \hat x \phi(r);
\qquad V_0 = g_\omega \omega(r); \qquad A_0 = 0 \cr
& \alpha \cdot V = {g_\rho \over 2}
\alpha\cdot( \hat x \wedge \tau ) \rho(r) \cr
& \alpha \cdot A = {g_\rho \over 2} \[
\alpha\cdot \tau \( A_S(r) -{1\over 3}A_T (r) \) +
(\alpha\cdot \hat x ) (\tau \cdot \hat x ) A_T (r) \] \cr }
\eqno(8.14)$$
If we introduce the hedgehog ansatz in the expressions for
energy (7.7) and the currents (7.9) we readily find the modifications
for the
total energy and the baryon, vector and axial currents. Their detailed
form can be seen at Appendix C.

Given a vector gauge invariant regularization of the fermion determinant
one should proceed as follows. The total energy is obtained by adding
the valence quark, the Dirac sea contribution (vector additively
regularized),
the bosonizing terms and the counterterms given by formula (8.9).
One should then minimize this
total energy with respect to arbitrary variations in the fields
$\sigma, \phi, \rho, A_S, A_T $ and $\omega$,
and determine the solution of the resulting equations of motion by some
iterative method until convergence is achieved. In the present case,
such a procedure requires many iterations [29, 30, 31]. In
addition, there is presently some concern [39,40,41] about the validity
of any of the schemes proposed so far [25,28,29,30,31]. It is clear,
however, that in either case the
proper anomalous structure has not been included since a vector additive
regularization has always been considered, i.e. the counterterms $\Delta
E$ are missing. Therefore, before embarking in a rather complicated
calculation we make an order of magnitude estimate treating the
counterterms as a small perturbation of the total energy. Clearly, such
a procedure can be justified a posteriori if the modifications
actually turn out to be small.

A quantitative estimate of the corrections can be obtained by
considering any of the self-consistent solutions available in the
literature [29,30,31], since in practice they do not differ too much
from
each other. On simple dimensional grounds and using the typical sizes of
the mesonic fields ($ \sim f_\pi $) and their typical extension
($ \sim 1 $ fm) for reasonable values of the parameters we estimate
the counterterm to the total energy to be of the order of few
MeV. A more systematic estimate using the self-consistent solutions
of ref. [29] gives values for $\Delta E < 10 $ MeV. This small number
does
not stem neither from a big cancellation among the different terms in the
expression for $\Delta E$ nor
from cancellations between the interior and the exterior of the soliton.
The size of the correction is to be compared to the typical
soliton energies $ E \sim 1500 $ MeV. Thus it seems more than reasonable
to treat the counterterms perturbatively. For the currents a similar
strategy can be considered. In this case we can compare the perturbative
contributions to the isoscalar nucleon radius and the axial coupling
constant to the self-consistent ones given by formulas (C.5), (C.6),
(C.7) and (C.10)
respectively, the difference being an indication of how far are we from
a self-consistent solution, i.e. they represent virial theorems. Again
the order of magnitude estimate predicts $ \Delta r
\sim 0.01 $ fm and $\Delta g_A^N \sim 0.01 $ in any calculation scheme,
much smaller than the typical values usually found ($r\sim 0.8$ fm and
$g_A^N \sim 0.5 $ [29,30,31]).
This result is also confirmed by an accurate calculation using the
solutions of ref. [29].
It is beyond
doubt that the modification of the Nambu--Jona-Lasinio model to
incorporate the proper QCD anomaly does not result in large changes
for the computed nucleon properties. At first sight this is a bit
surprising since in the meson sector $20\% $ deviations with respect
to the current algebra result have been obtained for
$\gamma\to\pi\pi\pi$ [23]. On the other
hand, the corrections induced by the counterterms involve vector mesons only
(see Appendix C) whose typical extension is $1/m_\rho$. Due to this
the corrections in the nucleon currents are of order $f_\pi/m_\rho\sim 0.1$,
i.e. they are small because of the large vector meson masses.
It is interesting to note here that similar trends have been
also found in the meson sector when calculating full momentum dependent
abnormal parity vertex functions [42].

{\bf 9. Effective action for Vector Mesons up to fourth order in
momenta}

In this section we further exploit our results to write down an
effective action for vector mesons and external fields up to fourth
order in momenta. The interesting point is that the anomaly does not
imply the coupling strength of strong decay processes like $\omega \to
3\pi $, $\omega \to \rho \pi $ or radiative vector meson decays like
$ \rho \to \gamma \pi $ etc., although the corresponding amplitudes
have abnormal pseudoparity (i.e. contain an
$\epsilon_{\mu\nu\alpha\beta}$ tensor).

 As we have said in section 6, up to fourth order the correct anomalous
action reproducing the proper QCD anomaly is given by
$$ \Gamma_{\rm WZ}^V [ U , v , a ; V , A ] =
\Gamma_{\rm WZ}^{\rm RL}[U,\A_R,\A_L ]
-\Gamma_{\rm ct}^{\rm RL}[v,V] - \Gamma_{\rm WZ}^{\rm RL} [ 1 , v_R ,
v_L ]
\eqno(9.1)$$
On the other hand the action
$$ \Gamma^V_{\rm WZ}  [U,v,a] = \Gamma_{\rm WZ}^{\rm RL}[U,v_R,v_L]
-  \Gamma_{\rm WZ}^{\rm RL} [ 1 , v_R , v_L ]
\eqno(9.2)$$
corresponding to $G_V=0$ and $\Delta\Gamma =0$ also reproduces the QCD
anomaly. Therefore both actions differ by
chirally covariant terms. A direct computation gives
$$\eqalign{\Gamma^V_{\rm WZ} [ U , v , a; V,A ] & = \Gamma^V_{\rm WZ}
 [U, v,a ] \cr & -
{iN_c \over 48\pi^2}\int \tr\Bigl\{ 2R \{ F_R , V_R \} + DV_R [ R , V_R]
-iR( R^2 +  V_R^2 ) V_R \cr & + {1\over 2} (RV_R)^2 + 2iU F_R V_R
U^\dagger
V_L + U R F_R U^\dagger V_L + i U V_R U^\dagger V_L F_L \cr &
+ UF_R R U^\dagger V_L + DV_R
RU^\dagger V_L U + i DV_R [ V_R , U^\dagger V_L U ] + U R^2 V_R
U^\dagger V_L \cr & + i
UV_R U^\dagger V_L L V_L + UV_R U^\dagger V_L^3 + {1\over 4} (U V_R
U^\dagger V_L )^2 \Bigr\} -{\rm p.c.} \cr }
\eqno(9.3)$$
where the following chirally covariant 1-forms
$$\eqalign{ &
R = U^\dagger \nabla U = U^\dagger d U
- i U^\dagger v^L U +i v^R \cr &
  L = U \nabla U^\dagger = U d U^\dagger
- i U v^R U^\dagger +i v^L \cr }
\eqno(9.4)$$
and 2-forms
$$\eqalign{ &
F_R = dv_R - iv_R^2 ; \quad F_L = dv_L - iv_L^2 \cr &
   DV_R = d V_R - i \{ V_R , v_R \} ; \quad
   DV_L = d V_L - i \{ V_L , v_L \} \cr }
\eqno(9.5)$$
have been introduced.
The transformation properties of these objects are
$$ \delta R = i[\epsilon_R,R] ; \qquad
   \delta F_R = i[\epsilon_R, F_R] ; \qquad
   \delta DV_R = i[\epsilon_R , DV_R ]
\eqno(9.6)$$
similarly for the left combinations.
It is important to mention that from the point of view of an effective
mesonic theory each term appearing in eq. (9.3) can have an arbitrary
coefficient since they are separately chirally invariant and do not
contribute to the anomaly equation. The coefficients in eq. (9.3)
represent the particular prediction of the Nambu--Jona-Lasinio model
with vector mesons so defined to satisfy the correct QCD anomaly.
Another
important point is that these terms describe low energy vector meson
strong and radiative abnormal parity decays although they are clearly
not anomalous.  In particular, if the external fields are set equal
to zero the total action is invariant under global chiral
transformations, and no internal anomaly appears. We will not attempt to
study the phenomenological
implications of the former action. Some of them have been considered in
ref. [43] for low momenta and in ref. [42] in a full momentum
dependent formalism, although the explicit form (9.3) has not been
given. Finally, as a partial check of our results, we consider the low
energy limit of the former abnormal parity non-anomalous action. This
corresponds to integrate vector mesons out at the mean field level or
equivalently in the large $N_c $ limit. In the lowest relevant order we
have the following equations of motion (see [35] and [29] for more
details)
$$
V_\mu^R = {i\over 2} (1-g_A^Q )R_\mu ; \qquad
V_\mu^L = {i\over 2} (1-g_A^Q )L_\mu
\eqno(9.7)$$
with $ g_A^Q $ the quark axial coupling constant [23,35]. Moreover
$$\eqalign{
DV_R &= {i\over 2}(1-g_A^Q ) \( -R^2 + iF_R - iU^\dagger F_L U \) \cr
DV_L &= {i\over 2}(1-g_A^Q ) \( -L^2 + iF_L - iU F_R U^\dagger \) \cr }
\eqno(9.8)$$
giving
$$ \Gamma^V_{\rm WZ} [ U, v, a ; V , A ]
= \Gamma^V_{\rm WZ} [U, v, a ] +  \cdots
\eqno(9.9)$$
where the dots denote terms of order sixth at least.
Therefore up to fourth order in the chiral expansion there are no
corrections to the effective abnormal parity action. This result is also
a consequence of the uniqueness of the counterterms to the Wess-Zumino
action (see the discussion at the end of section 6).

\vskip1cm
{\bf 10. Conclusions and Summary }

In the present work we have investigated the anomalous sector
of the Nambu--Jona-Lasinio model with vector mesons and its possible
implications on the properties of the nucleon described as a system of
three bound valence quarks in a self-consistent solitonic background of
$\sigma$, $\pi$,
$\rho$, $A$ and $\omega$ mesons. In most cases, calculations within this
model have ignored the fact that it does not reproduce the correct
QCD chiral anomaly. For the model to do so
for non-vanishing vector meson fields
it is necessary to modify the
usual definition of the fermionic determinant. This can be done by
subtracting
suitable local and polynomial counterterms to the action in the vector
dynamical and external fields.
This represents another solution  which was overlooked in [23],
hence the conclusion there, that the NJL model with vector mesons cannot
reproduce the correct QCD anomaly, is incorrect.
These counterterms only modify the
abnormal parity vertices at zero momentum transfer and in the chiral
limit, and hence leave many meson properties such as meson propagators
and the momentum dependence of mesonic form factors unaffected.
However, there appear abnormal parity modifications
appear in the Current-Field identities at leading order in large $N_c$.
In the soliton sector clear corrections appear in the nucleon mass and
the axial and vector currents. For hedgehog profiles we have evaluated
the numerical corrections to the soliton energy, the isoscalar nucleon
radius and the axial coupling constant induced by the counterterms. We
have found that they account for less than $1\%$ of the total
magnitude of
the computed observables. As the counterterms only involve vector
mesonic fields, and the corrections take place at zero momentum in the
amplitudes, they are mainly sensitive to the tail of the vector
meson fields only. The smallness of the counterterms in the nucleon
might be understood due to the high vector meson masses. We conclude
that the fact that the previous regularizations of the
Nambu--Jona-Lasinio model
do not fulfill the proper QCD chiral anomaly does not have practical
dramatic consequences in the previous solitonic calculations.

\vskip1cm
{\bf ACKNOWLEDGEMENTS}

{\sl We thank F. D\"oring for numerical help and K. Goeke for
encouragement. One of us (E.R.A.) acknowledges NIKHEF for hospitality.

This work has been partially supported by the DGICYT under contract
PB92-0927 and the Junta de Andaluc{\'{\i}}a (Spain) as well as FOM and
NWO (The Netherlands).}

\newpage

{\bf Appendix A. Chiral Transformations of the Dirac operator in
Minkowski space}

We define the effect of local chiral transformations
on the Dirac operator, both in vector-axial notation as well as in
right-left notation in Minkowski space. We follow Bjorken-Drell [44]
(Itzykson-Zuber [45]) notation for the gamma matrices throughout.

{\sl A.1 Vector-Axial Notation}

The Dirac operator $\D$ in Minkowski space is taken to be
$$
i \D
= i \slashchar\partial + {\slashchar{\cal V}} + {\slashchar{\cal A}}
\gamma_5 - ( {\cal S}+ i \gamma_5 {\cal P} )
\eqno({\rm A}.1)$$
with $S$,$P$,$V_\mu$,$A_\mu$ hermitean flavour fields
$ S = {1\over 2}\sum_{a=0}^{N_F^2 -1} \lambda_a S^a (x) , \cdots $,
with the flavour matrices $\lambda_a $ normalized as $\tr ( \lambda_a
\lambda_b ) = 2 \delta_{ab}$
and $\lambda_a \lambda_b = ( d_{abc} +
i\, f_{abc} )\lambda_c $ with $a,b,c=0, \dots, N_F^2 -1 $.
Under chiral (vector and axial)  \underbar{local} transformations the
Dirac operator transforms as
$$
 \D \to \D^g = e^{+i\epsilon_V (x) -i\epsilon_A (x) \gamma_5 } \D
e^{-i\epsilon_V (x)-i\epsilon_A (x) \gamma_5 }
\eqno({\rm A}.2)$$
with
$$
\epsilon_V (x) = \sum_a  \epsilon_V^a (x) \lambda_a ; \qquad
\epsilon_A (x) = \sum_a  \epsilon_A^a (x) \lambda_a
\eqno({\rm A}.3)$$
The induced transformations on the fields are given by
$$ i\D \to i\D^g = i \slashchar\partial + {\slashchar{\cal V}}^g +
{\slashchar{\cal A}}^g \gamma_5 - ( {\cal S}^g + i \gamma_5 {\cal P}^g )
\eqno({\rm A}.4) $$
In the infinitesimal case this is equivalent to
$$\eqalign{
& \delta \V^\mu = [ {\cal D}^\mu , \epsilon_V ] +
                i [ \epsilon_A , \A^\mu ] \cr
& \delta \A^\mu = i [ \epsilon_V , \A^\mu ]
                  + [ {\cal D}^\mu , \epsilon_A ] \cr
& \delta \S = i [ \epsilon_V , \S ]
             + \{ \epsilon_A , \P \} \cr
& \delta \P = i [ \epsilon_V , \P ]
             - \{ \epsilon_A , \S \} \cr }
\eqno({\rm A}.5)$$
where the vector covariant derivative reads
$${\cal D}_\mu = \partial_\mu -i\V_\mu
\eqno({\rm A}.6)$$
In addition, the dynamical and external quark fields satisfy the
following transformation properties
$$\eqalign{
&\delta q = i( \epsilon_V +  \epsilon_A \gamma_5 ) q \cr
&\delta\bar q = -i\bar q (\epsilon_V - \epsilon_A \gamma_5 ) \cr
&\delta\eta = i( \epsilon_V -  \epsilon_A \gamma_5 ) \eta \cr
&\delta\bar \eta =-i\bar \eta (\epsilon_V + \epsilon_A \gamma_5 )\cr }
\eqno({\rm A}.7)$$
defined to make the bilinear forms $\bar q \D q $ and $\bar \eta \D^{-1}
\eta $ invariant under chiral local transformations.

{\sl A.2 Left-Right Notation}

We rewrite the Dirac operator as
$$ \D = \D_R P_R + \D_L P_L
\eqno({\rm A}.8)$$
with the projection operators on chirality
$$
P_R = {1\over2} ( 1+ \gamma_5 ) ; \qquad
P_L = {1\over2} ( 1- \gamma_5 )
\eqno({\rm A}.9)$$
The right and left Dirac operators are given by
$$
i \D_R = i \slashchar\partial + {\slashchar \A}_R - \M ; \qquad
i \D_L = i \slashchar\partial + {\slashchar \A}_L - \M^\dagger
\eqno({\rm A}.10)$$
with
$$\eqalign{
& \M       = \S + i\P  ;  \qquad  \M^\dagger =  \S - i\P  ;  \cr
& \A^\mu_R = \V^\mu + \A^\mu ; \qquad  \A^\mu_L = \V^\mu - \A^\mu \cr }
\eqno({\rm A}.11)$$
then we have the following transformation properties under infinitesimal
chiral rotations
$$\eqalign{
& \delta \M = i\epsilon_L \M - i \M \epsilon_R \cr
& \delta \M^\dagger = i\epsilon_R \M^\dagger-i\M^\dagger \epsilon_L  \cr
& \delta \A^\mu_R = \partial^\mu \epsilon_R + i[ \epsilon_R , \A^\mu_R ]
= [ {\cal D}^\mu_R , \epsilon_R ] \cr
& \delta \A^\mu_L = \partial^\mu \epsilon_L + i[ \epsilon_L , \A^\mu_L ]
= [ {\cal D}^\mu_L , \epsilon_L ] \cr }
\eqno({\rm A}.12)$$
where
$$
\epsilon_R = \epsilon_V + \epsilon_A ; \qquad
\epsilon_L = \epsilon_V - \epsilon_A ; \qquad
{\cal D}_\mu^R = \partial_\mu -i\A_\mu^R ; \qquad
{\cal D}_\mu^L = \partial_\mu -i\A_\mu^L
\eqno({\rm A}.13)$$
have been defined. Finally, the left and right currents are given by
$$
J_\mu^{R,L} = J_\mu^V \pm J_\mu^A
\eqno({\rm A}.14)$$

\newpage

{\bf Appendix B. Effective action, regularization and anomaly in
Euclidean space}

For completeness we give below our particular conventions for Euclidean
space (hatted quantities)
$$ \hat x^0 = i x^0 ; \qquad \hat x^i = x^i ;
\qquad \hat x^\mu = \hat x_\mu = ( \hat x^0 , \hat x^i )
\eqno({\rm B}.1)$$
$$ \hat V^0 ( \hat x^0 , \hat x^i ) = i V^0 ( x^0 , x^i ) ; \qquad
   \hat V^i ( \hat x^0 , \hat x^i ) =   V^i ( x^0 , x^i ) ; \qquad
   \hat V^\mu = \hat V_\mu = ( \hat V^0 , \hat V^i )
\eqno({\rm B}.2)$$
$$ \hat \partial_0 = -i\partial_0 ; \qquad
\hat \partial^i = + \partial_i = -\partial^i ; \qquad
\hat \partial^\mu = \hat \partial_\mu = {\partial \over \partial
\hat x^\mu}
\eqno({\rm B}.3)$$
$$ \hat \gamma^0 = -i \gamma^0 ; \qquad \hat \gamma^i = -\gamma^i ;
\qquad \hat \gamma^\mu = \hat \gamma_\mu = ( \hat \gamma^0 , \hat
\gamma^i )= - \hat \gamma_\mu^\dagger
\eqno({\rm B}.4)$$
$$ \hat x \cdot \hat y  = \hat x^\mu \hat y_\mu =
-x^\mu y_\mu = - x \cdot y
\eqno({\rm B}.5)$$
$$ \hat \gamma_5 = \hat \gamma^0
   \hat \gamma^1  \hat \gamma^2 \hat \gamma^3 =
   i \gamma^0 \gamma^1  \gamma^2 \gamma^3 = \gamma_5
\eqno({\rm B}.6)$$
$$ \slashchar{\hat \partial} = \hat \gamma \cdot \hat \partial =
-\gamma \cdot \partial = -\gamma^\mu \partial_\mu =
-\slashchar{\partial}
; \qquad \slashchar{\hat V} = \hat \gamma \cdot \hat V =  \gamma \cdot
V = \slashchar{ V}
\eqno({\rm B}.7)$$
$$\hat \epsilon^{0123} = \hat \epsilon_{0123}= + 1 ; \qquad
\epsilon^{0123} = -\epsilon_{0123} = + 1
\eqno({\rm B}.8)$$
$$  \hat \epsilon^{\mu\nu\alpha\beta} \hat A^\mu \hat B^\nu
\hat C^\alpha \hat D^\beta =
- i \epsilon_{\mu\nu\alpha\beta} A^\mu B^\nu C^\alpha D^\beta
\eqno({\rm B}.9)$$
The factor in the exponential reads
$$ \exp\Bigl\{ i \int d^4 x \bar q i\D q \Bigr\} =
   \exp\Bigl\{ - \int d^4 \hat x \bar q i \hat \D q \Bigr\}
\eqno({\rm B}.10)$$
with the Dirac operator in Euclidean space
$$ i \hat \D = i \slashchar{\hat \partial} - {\slashchar {\hat{\cal
V}}} - {\slashchar{\hat{\cal A}}}\hat \gamma_5 +  \hat{\cal S}+ i
\hat \gamma_5
\hat{\cal P}
\eqno({\rm B}.11)$$
whose hermitean conjugate is given by
$$ -i \hat \D^\dagger = - i \slashchar{\hat \partial} + {\slashchar
{\hat{\cal V}}} - {\slashchar{\hat{\cal A}}}\hat \gamma_5
+  \hat{\cal S}- i \hat \gamma_5 \hat{\cal P}
\eqno({\rm B}.12)$$
Notice that the hermitean conjugation in Euclidean space $\hat \D \to
\hat \D^\dagger $ corresponds to the operation $\D \to \D_5 $
 in Minkowski space (see eq. (4.2)).  The fermion determinant is
$$ \overline{\Det} (i\hat \D) =
\exp(-\hat{\overline{W}}) = \overline{\Det}(i \D)= \exp(i
\overline{W})
\eqno({\rm B}.13)$$
The vector-additively regularized real part of the fermionic
contribution to the effective action reads then
$$ {\rm Re}\, \hat{\overline{W}}= {1\over 4} {\rm Sp} \int_0^\infty
{d\tau \over \tau} \Phi (\tau ) \left[e^{-\tau \hat \D
\hat\D^\dagger}+e^{-\tau \hat \D^\dagger \hat \D} \right]
\eqno({\rm B}.14)$$
Here $\Phi(\tau )$ is a generalized proper-time regularization
function with $\Phi(0 )= \Phi'(0)=0$ and
$\Phi(\infty)=1$. The Pauli-Villars regularizations correspond to the
choice $ \Phi = \sum_i c_i \exp (-\tau \Lambda_i^2) $.
The imaginary part takes the form
$$ {\rm Im }\, \hat{\overline{W}}= {i\over 2} {\rm Sp} \left[ \log
(i \hat \D ) - \log(-i \hat \D^\dagger ) \right]
\eqno({\rm B}.15)$$
The variation of the determinant in such regularization involves the
imaginary part of the Euclidean action only giving
$$\eqalign{ \delta \log \overline{\Det}
 ( i \hat \D ) & = - i \delta {\rm Im}
\hat{\overline{W}} = \cr & -{iN_c \over 4\pi^2}
\hat \epsilon_{\mu\nu\alpha\beta} \int d^4 \hat x \tr \Bigl\{ \epsilon_A
(x) \[ {1\over 4} [ i \hat {\cal D}_\mu , i\hat {\cal D}_\nu ]
              [ i \hat {\cal D}_\alpha , i\hat {\cal D}_\beta ]
-{1\over 3} \hat \A_\mu \hat \A_\nu \hat \A_\alpha \hat \A_\beta \cr &
-{2\over 3} \hat \A_\mu [ i \hat {\cal D}_\nu, i \hat {\cal D}_\alpha ]
\hat \A_\beta
-{1\over 6} \{ [ i\hat {\cal D}_\mu , i\hat {\cal D}_\nu ] ,
\hat \A_\alpha \hat \A_\beta \}
+{1\over 3} [ i\hat {\cal D}_\mu , \hat \A_\nu ]
[ i\hat {\cal D}_\alpha, \hat \A_\beta ] \] \Bigr\} \cr }
\eqno({\rm B}.16)$$
where $ i\hat {\cal D}_\mu = i\hat \partial_\mu - \hat \V_\mu $ has
been defined.

\newpage
{\bf Appendix C. Results for Hedgehog Profiles}

Here are the expressions for the modification of the computed quantities
and the total results for self-consistent fields not given in the main
text.

{\sl C.1 Modification of nucleon observables}

\underbar{Mean field energy}
$$\eqalign{ \Delta E [ \omega, \rho , A ] & = -{N_c \over 12\pi^2}
g_\rho^2 g_\omega \int_0^\infty 4\pi r^2 dr \omega
\Bigl\{ 2 \rho ( A_S' - {1\over 3} A_T' + {1\over r} A_T ) \cr &
+4 (\rho' + {2\over r} \rho ) ( A_S - {1\over 3} A_T )
+ {3 g_\rho \over 4} (A_S + {2\over 3} A_T )
\[ ( A_S - {1\over 3} A_T )^2 + \rho^2 \]   \Bigr\} \cr }
\eqno({\rm C}.1)$$

\underbar{Isoscalar baryon density }
$$ \Delta J^B_0 = {N_c\over 4\pi^2} g_\rho^2 {1\over r^2} {d\over dr}
\Bigl\{ r^2 (A_S-{1\over 3} A_T ) \rho \Bigr\}
\eqno({\rm C}.2)$$

\underbar{Vector-isovector current }
$$\eqalign{ & (\Delta J^V )^i_a = -\epsilon_{iak} \hat x_k {N_c \over
24\pi^2} g_\rho g_\omega \Bigl\{ 3\[(A_S -{1\over 3} A_T)\omega\]' -
{3 A_T \omega \over r} - 2 g_\rho \omega \rho (A_S + {2\over 3} A_T )
\Bigl\} \cr }
\eqno({\rm C}.3)$$

\underbar{Axial isovector current }
$$\eqalign{ (\Delta J^A)^i_a = {N_c\over 24\pi^2} g_\rho g_\omega &
\Bigl\{ \delta_{ia} \[ 3\omega'\rho - 3\omega(\rho' + {\rho\over r})
-2g_\rho\omega(A_S -{1\over 3}A_T) (A_S + {2\over 3}A_T )\] \cr &
+  \hat x_i \hat x_a  \[ -3\omega'\rho + 3\omega(\rho' -
{\rho\over r}) - 2g_\rho\omega \( \rho^2 - (A_S -{1\over 3}A_T) A_T \)
\] \Bigr\} \cr }
\eqno({\rm C}.4)$$

\underbar{Isoscalar radius }
$$\langle \Delta r^2 \rangle_N^{I=0} = -{N_c\over 2\pi^2}\int_0^\infty
4\pi r^3 dr (A_S-{1\over 3} A_T ) \rho
\eqno({\rm C}.5)$$

\underbar{Axial coupling constant }
$$ \Delta g_A^N = - {N_c\over 24\pi^2} g_\rho g_\omega {2\over 3}
\int_0^\infty 4\pi r^2 dr 2\omega \[ 2 ( \rho' + {2\rho \over r})
+ {1\over 3 } g_\rho \rho^2 + g_\rho (A_S -{1\over 3} A_T )
(A_S + {1\over 3} A_T ) \]
\eqno({\rm C}.6)$$

{\sl C.2 Self-consistent nucleon observables}

\underbar{Isoscalar baryon density }
$$\eqalign{ (J^B )^0 = -{m_\omega^2 \over g_\omega} \omega
+{N_c\over 12\pi^2} g_\rho^2 & \Bigl\{ \rho \( A_S'-{1\over 3} A'_T
-{2 A_T \over r} \) - (A_S -{1\over 3} A_T )(\rho' + {2\over r} \rho)
\cr & -{3\over 4} g_\rho (A_S + {2\over 3}A_T)\[ (A_S -{1\over 3}A_T)^2
+ \rho^2 \] \Bigr\} \cr }
\eqno({\rm C}.7)$$

\underbar{Isospin current }
$$\eqalign{ ( J^V )^i_a = \epsilon_{iak} \hat x_k & \Bigl\{
{m_\rho^2 \over g_\rho} \rho + {N_c \over 24\pi^2} g_\rho g_\omega
\[ (A_S -{1\over 3} A_T)\omega' \cr &
-(A_S ' -{1\over 3} A_T ')\omega + {A_T \omega \over r} +
 {1\over 2} g_\rho \omega \rho (A_S + {2\over 3} A_T )  \] \Bigl\} \cr }
\eqno({\rm C}.8) $$

\underbar{Axial current}
$$\eqalign{ ( J^A )^i_a & = -\delta_{ia} \Bigl\{ {m_\rho^2 \over g_\rho}
( A_S - {A_T \over 3} ) + {N_c \over 24\pi^2} g_\rho g_\omega \[
-\omega'\rho + \omega \rho' + {\rho\omega\over r} + {1\over 2} g_\rho
\omega (A_S - {A_T \over 3} ) (A_S + {2 A_T \over 3}) \Bigr\} \cr &
-\hat x_i \hat x_a \Bigl\{ {m_\rho^2 \over g_\rho}  A_T
+ {N_c \over 24\pi^2}
g_\rho g_\omega \[ \omega'\rho - \omega \rho' + {\rho\omega\over r} -
{1\over 2} g_\rho \omega \( (A_S - {1\over 3} A_T ) A_T  - \rho^2 \)\]
\Bigr\} \cr }
\eqno({\rm C}.9)$$

\underbar{Axial coupling constant}

$$g_A^N = -{2\over 3} \int d^3 x \Bigl\{ {m_\rho^2 \over
g_\rho} A_S
+{N_c \over 36\pi^2} g_\rho g_\omega \[ -\omega' \rho + \omega\rho' +
{2\omega\rho\over r} + {3\over 4} g_\rho( A_S^2 -{1\over 9} A_T^2
+{1\over 3} \rho^2 ) \] \Bigr\}
\eqno({\rm C}.10)$$

\newpage
{\bf Appendix D. Five dimensional expression for the gauged
Wess-Zumino term}

In this appendix we give an alternative form for the gauged Wess-Zumino
term both in its right-left and vector versions, $\Gamma^{\rm
RL}_{\rm WZ}(U,v_R,v_L)$ and $\Gamma^V_{\rm WZ}(U,v,a)$
respectively, as five-dimensional integrals.
The interesting feature of these formulas is that, oposite to
the more conventional form, (see eq. (6.1)), the chiral transformation
properties become more evident.

The topological Wess-Zumino term reads [4]
$$\Gamma_{\rm WZ}[U] =  -{iN_c \over 48\pi^2} \int_{D_5} \tr  \[
{1\over 10} U_R^5 \] \eqno({\rm D}.1)$$
where $D_5$ represents a five-dimensional manifold with
boundary the compactified four-dimensional space-time,
$U_R = U^\dagger dU$ is a 1-form living in $D_5$ and $U(x,s)$ is
an interpolating unitary flavour field, $U^\dagger U=1$, satisfying
$U(x,1)=U(x)$ and $U(x,0)$ a constant matrix.
The following two observations will be used below.
Since the integrand is a closed
form (i.e. locally exact) the result does not depend (modulo homotopy classes)
on the particular choice of $D_5$. Therefore the action is well-defined
modulo $2\pi i$.
In addition, in this form the topological Wess-Zumino term is manifestly
invariant under global chiral transformations but not under
local ones. As pointed out in ref. [4] the traditional minimal substitution
method cannot be applied in its standard form since the topological
action is non local.
In fact the existence of the anomaly prevents a full chiral gauging.
Thus one has to resort to other methods [38].
For instance, eq. (6.1) can be obtained by trial and error gauging.

An alternative way to obtain the gauged Wess-Zumino term is the following.
We consider the minimal substitution rule directly in five
dimensions, i.e. we make $U_R\to R$ with $R$ the natural five-dimensional
extension of eq. (9.4). This requires introducing
five-dimensional gauge fields as well.
Clearly $R(x,s)$ is chirally covariant
and hence the new action is formally chirally invariant.
However, since the integrand $R^5$ is not a closed 5-form the
action is no longer independent of the choice of $D_5$ modulo
$2\pi i$. To reestablish the one-valuedness of the action one
has to supplement the integrand with additional terms satisfying
the following two conditions: i) the total sum has to be closed
and ii) the new terms have to vanish in the absence of gauge fields.
This uniquely determines the result. For the right-left
representation we find
$$\eqalign{
\Gamma_{\rm WZ}^{\rm RL} [ U, v_R , v_L ] = -{iN_c \over 48\pi^2
} \int_{D_5} \tr  \[ &
{1\over 10} R^5 -
2R F_R^2 + iR^3 F_R -R F_R U^\dagger F_L U \cr + & 2iv_R F_R^2 -
v_R^3 F_R -{i\over 5} v_R^5 \] - {\rm p.c.} \cr }
\eqno({\rm D}.2)$$
One should mention that the usual integration by parts is
precluded in this formula since $D_5$ has a boundary.
One can check that the difference between the
gauged and the non gauged Wess-Zumino terms is in fact the five-dimensional
integral of an exact 5-form which is the differential of the corresponding
four-dimensional piece in (6.1).
On the other hand, the vector representation,
again obtained by Bardeen's subtraction, reads
$$\eqalign{
\Gamma_{\rm WZ}^{\rm VA} [ U, v, a] = -{iN_c \over 48\pi^2}
\int_{D_5} \tr \[ & {1\over 10} R^5 -
2R F_R^2 + iR^3 F_R -R F_R U^\dagger F_L U \cr + &
2iaF_R F_L +
4iaF_R^2 - 8a^3 F_R -i{16\over 5} a^5 \] - {\rm p.c.} \cr }
\eqno({\rm D}.3)$$

Besides the fact that the last two expressions are more compact than
the corresponding ones, eqs. (6.1) and (6.6), the chirally
breaking terms are manifestly polynomial and hence the anomaly.
Let us remark that these polynomial terms do not form an exact differential
in five-dimensions and hence they cannot be subtracted by adding
four-dimensional polynomial counterterms. In other words the anomaly
cannot be removed.

{\bf Appendix E. CP-violating currents}

As already mentioned in section 6, the counterterms are unique if
CP-invariance is invoked.
The only possible CP-violating counterterms are given by
$$ \Gamma_{\rm ct}^{\rm CP} = -i c {N_c \over 48\pi^2} \int
\tr\[ V_R^2 F_R - V_L^2 F_L \]
\eqno({\rm E}.1)$$
with $c$ an arbitrary real constant as implied by CPT.
The contributions to the currents are
$$\int \tr\( v J_V^{\rm CP} + a J_A^{\rm CP} \)= -i c {N_c \over
24\pi^2} \int \tr\( dv(VA+AV) + da(V^2 + A^2) \)
\eqno({\rm E}.2)$$
In the two flavour case we have
$$ \eqalign{ ( J^B)^{\rm CP}_\mu & = 0 \cr
          ( \vec J^V)^{\rm CP}_\mu & = { N_c \over 48\pi^2 } c g_\rho^2
\epsilon_{\mu\nu\alpha\beta} \partial^\nu ( \vec \rho^\alpha \wedge
\vec A^\beta ) \cr
          ( \vec J^A)^{\rm CP}_\mu & = { N_c \over 96\pi^2 } c g_\rho^2
\epsilon_{\mu\nu\alpha\beta} \partial^\nu ( \vec \rho^\alpha \wedge
\vec \rho^\beta + \vec A^\alpha \wedge \vec A^\beta ) \cr }
\eqno({\rm E}.3)$$
For hedgehog profiles all CP-violating currents vanish except the time
component of the axial current
$$
( \vec J_A )_0 = { N_c \over 24\pi^2} c g_\rho^2 \hat x_a
\Bigl\{ \rho(\rho'+ {\rho\over r} ) +
(A_S -{A_T \over 3})( A_S' -{A_T'\over 3} - {A_T \over r} ) \Bigr\}
\eqno({\rm E}.4)$$

\vfill\eject

\centerline{\bf REFERENCES}

\item{[1]} {W. Bardeen, Phys. Rev. {\bf 184} (1969) 1848.}

\item{[2]} {S. L. Adler and W. A. Bardeen, Phys. Rev. {\bf 182}
(1969) 1517.}

\item{[3]} {J. Wess and B. Zumino,  Phys. Lett. {\bf B 37} (1971)
95.}

\item{[4]} {E. Witten, Nucl. Phys. {\bf B 223} (1983) 433.}

\item{[5]} {Y. Brihaye, N. K. Pak and P. Rossi; Nucl. Phys. {\bf B
254}  (1985) 71.}

\item{[6]} {T. Fujiwara, T. Kugo, H. Terao, S. Uehara and K. Yamawaki,
Prog. Theor. Phys. {\bf 73} (1985) 926.}

\item{[7]} { For reviews see e.g. U. G. Mei\ss ner, Phys. Rep. {\bf
161}
(1988) 213; M. Bando, T. Kugo, and K. Yamawaki, Phys. Rep. {\bf 164}
(1988) 217 and references therein. }

\item{[8]} {K. Fujikawa, Phys. Rev. Lett. {\bf 44} (1980) 1733;
                         Phys. Rev. {\bf D 21} (1980) 2848.}

\item{[9]} {A. P. Balachandran, G. Marmo, V. P. Nair and C. G.
Trahern,
                         Phys. Rev. {\bf D 25} (1982) 2713.}

\item{[10]} {Y. Nambu and G. Jona-Lasinio, Phys. Rev. {\bf 122} (1961)
                    345; {\bf 124} (1961) 246.}

\item{[11]} {For reviews see e.g.
U. Vogl and W. Weise, {\it Progress in Particle and Nuclear Physics}
                 vol. {\bf 27} (1991) 195;
 S. P. Klevansky, Rev. Mod. Phys. {\bf 64} (1992) 649;
M. K. Volkov, Part. and Nuclei  {\bf B 24} (1993) 1;
 Th. Mei\ss ner, E. Ruiz Arriola, A. Blotz and K. Goeke, Rep. Prog.
Phys. (to appear) ; T. Hatsuda and T. Kunihiro, Phys. Rep. {\bf 247}
(1994)221.}

\item{[12]}  {A. Dhar and R. S. Wadia, Phys. Rev. Lett.
               {\bf 52} (1984) 959.}

\item{[13]} {A. Dhar, R. Shankar and S. R. Wadia,
                       Phys. Rev. {\bf D 31} (1985) 3256.}

\item{[14]} {D. Ebert and H. Reinhardt,
                        Nucl. Phys. {\bf B 271} (1986) 188.}

\item{[15]} {M. Wakamatsu and W. Weise,
                         Z. Phys. {\bf A 331} (1988) 173.}

\item{[16]} {M. Wakamatsu,  Ann. Phys. {\bf 193} (1989) 287.}

\item{[17]}  {M. Jaminon, R. Mendez-Galain, G. Ripka, P. Stassart,
                        Nucl. Phys. {\bf A 537} (1992) 418.}

\item{[18]} {V. Bernard, R. L. Jaffe and U. G. Mei\ss ner,
                        Nucl. Phys. {\bf B 308} (1988) 753.}

\item{[19]} {V. Bernard and U. G. Mei\ss ner,
                        Nucl. Phys. {\bf A 489} (1988) 647.}

\item{[20]}  {S. Klimt, M. Lutz, U. Vogl and W. Weise,
                        Nucl. Phys. {\bf A 516} (1990) 429.}

\item{[21]}  {M. Takizawa, K. Tsushima, Y. Kohyama and K. Kubodera,
                        Nucl. Phys. {\bf A 507} (1990) 611.}

\item{[22]} {N. M. Kroll, T. D. Lee and B. Zumino, Phys. Rev. {\bf
157} (1967) 137.}

\item{[23]} {E. Ruiz Arriola and L.L. Salcedo
                        Phys. Lett. {\bf B 316} (1993) 148.}

\item{[24]}   {V. Bernard, A.H. Blin, B. Hiller, U.G. Mei\ss ner and
            M.C. Ruivo, Phys. Lett. {\bf B 305} (1993) 163.}

\item{[25]}   {C. Sch\"uren, E. Ruiz Arriola and K. Goeke,
                       Phys. Lett. {\bf B 287} (1992) 283.}

\item{[26]}   {R. Alkofer, H. Reinhardt, H. Weigel and U. Z\"uckert,
                               Phys. Rev. Lett. {\bf 69} (1992) 1874}

\item{[27]} {F. D\"oring, E. Ruiz Arriola and K. Goeke,
                        Z. Phys. {\bf A 344} (1992) 159.}

\item{[28]}   {R. Alkofer, H. Reinhardt, H. Weigel and U. Z\"uckert,
                               Phys. Lett. {\bf B 298} (1993) 132.}

\item{[29]} {C. Sch\"uren, F. D\"oring, E. Ruiz Arriola and K. Goeke,
                       Nucl. Phys. {\bf A 565} (1993) 687.}

\item{[30]} {E. Ruiz Arriola, C. Sch\"uren, F. D\"oring  and K. Goeke,
                       J. Phys. {\bf G 20} (1994) 399.}

\item{[31]}   {U. Z\"uckert, R. Alkofer, H. Reinhardt and H. Weigel,
             Nucl. Phys. {\bf A 570} (1994) 445.}

\item{[32]} {J. Bijnens and J. Prades,
                       Phys. Lett. {\bf B 320} (1994) 201.}

\item{[33]} {W. Bardeen and B. Zumino,
                      Nucl. Phys. {\bf B 244} (1984) 421.}

\item{[34]} {\"O. Kaymak{\c c}alan, S. Rajeev and J. Schechter,
                      Phys. Rev. {\bf D 30} (1984) 594.}

\item{[35]} {E. Ruiz Arriola, Phys. Lett. {\bf B 253} (1991) 430.}

\item{[36]} {T. Eguchi, Phys. Rev. {\bf D 14} (1975) 2755.}

\item{[37]} {W. Pauli and F. Villars, Rev. Mod. Phys. {\bf 21} (1949)
434. }

\item{[38]} {R. D. Ball, Phys. Rep.{ \bf 182} (1989) 1. }

\item{[39]} {H. Weigel, U. Z\"uckert, R. Alkofer and H. Reinhardt,
Nucl. Phys. {\bf A 585} (1995) 513. }

\item{[40]} {C. Sch\"uren, PhD Thesis, Ruhr-Universit\"at Bochum, June
1994.}

\item{[41]} {F. D\"oring, C. Sch\"uren, E. Ruiz Arriola, T. Watabe and
K. Goeke, Granada preprint, February 1995. UG-DFM-2/95.}

\item{[42]} {J. Bijnens and J. Prades,
                       Nordita preprint ; NORDITA-94/11 N,P }

\item{[43]} {J. Prades,
                       Z. Phys {\bf C 63} (1994)491.}

\item{[44]} {J. D. Bjorken and S. D. Drell, {\it Relativistic Quantum
Mechanics}, McGraw-Hill, New York(1972).}

\item{[45]} {C. Itzykson and J. B. Zuber, {\it Quantum Field Theory}
McGraw-Hill, New York (1985).}

\bye